\documentclass[11pt,a4paper]{article}
\pdfoutput=1
\usepackage[english]{babel}
\usepackage{graphicx}
\usepackage{authblk}
\usepackage[maxnames=3,style=numeric,citestyle=numeric-comp,url=false,doi=false,backend=bibtex]{biblatex}

\DeclareBibliographyDriver{article}{%
  \usebibmacro{bibindex}%
  \usebibmacro{begentry}%
  \usebibmacro{author/translator+others}%
  \setunit{\printdelim{nametitledelim}}\newblock
  \newunit
  \printlist{language}%
  \newunit\newblock
  \usebibmacro{byauthor}%
  \newunit\newblock
  \usebibmacro{bytranslator+others}%
  \newunit\newblock
  \printfield{version}%
  \newunit\newblock
  \usebibmacro{journal+issuetitle}%
  \newunit
  \usebibmacro{byeditor+others}%
  \newunit
  \usebibmacro{note+pages}%
  \newunit\newblock
  \usebibmacro{finentry}}

\DeclareBibliographyDriver{inproceedings}{%
  \usebibmacro{bibindex}%
  \usebibmacro{begentry}%
  \usebibmacro{author/translator+others}%
  \setunit{\printdelim{nametitledelim}}\newblock
  \newunit
  \printlist{language}%
  \newunit\newblock
  \usebibmacro{byauthor}%
  \newunit\newblock
  \usebibmacro{bytranslator+others}%
  \newunit\newblock
  \printfield{version}%
  \newunit\newblock
  \usebibmacro{journal+issuetitle}%
  \newunit
  \usebibmacro{byeditor+others}%
  \newunit
  \usebibmacro{note+pages}%
  \newunit\newblock
  \usebibmacro{finentry}}

\AtEveryBibitem{\clearfield{month}}

\begin{document}

\selectlanguage{\english}
\title{\vspace{-2.0 cm} \bf Experimental nuclear astrophysics in Italy}

\author[1]{C.~Broggini\thanks{carlo.broggini@pd.infn.it}}
\author[2,3]{O.~Straniero}
\author[4,5]{M.G.F.~Taiuti}
\author[6]{G.~de Angelis}
\author[7]{G.~Benzoni}
\author[8,9]{G.E.~Bruno}
\author[10,11]{S.~Bufalino}
\author[12]{G.~Cardella}
\author[9]{N.~Colonna}
\author[13]{M.~Contalbrigo}
\author[14]{G.~Cosentino}
\author[2,15]{S.~Cristallo}
\author[16]{C.~Curceanu}
\author[12]{E.~De~Filippo}
\author[1]{R.~Depalo}
\author[17,18]{A.~Di~Leva}
\author[11]{A.~Feliciello}
\author[14]{S.~Gammino}
\author[6]{A.~Galat\`a}
\author[14]{M.~La Cognata}
\author[19,20]{R.~Lea}
\author[7,21]{S.~Leoni}
\author[12]{I.~Lombardo}
\author[9]{V.~Manzari}
\author[14]{D.~Mascali}
\author[22,23]{C.~Massimi}
\author[24]{A.~Mengoni}
\author[1,25]{D.~Mengoni}
\author[6]{D.R.~Napoli}
\author[15,26]{S.~Palmerini}
\author[20]{S.~Piano}
\author[12]{S.~Pirrone}
\author[14]{R.G.~Pizzone}
\author[12,27]{G.~Politi}
\author[4,5]{P.~Prati}
\author[6]{G.~Prete}
\author[14]{P.~Russotto}
\author[9]{G.~Tagliente}
\author[26]{G. M.~Urciuoli}

\affil[1] {INFN - Sezione di Padova, Via Marzolo 8, I-35131, Padova (Italy)}
\affil[2] {INAF - Osservatorio Astronomico d'Abruzzo, Via Mentore Maggini, I-64100 Teramo (Italy)}
\affil[3] {INFN - Laboratori Nazionali del Gran Sasso, Via G. Acitelli 22, I-67100 L'Aquila (Italy)}
\affil[4] {Universit\`a di Genova, Dipartimento di Fisica, Via Dodecaneso 33, I-16146 Genova (Italy)}
\affil[5] {INFN - Sezione di Genova, Via Dodecaneso 33, I-16146, Genova (Italy)}
\affil[6] {INFN - Laboratori Nazionali di Legnaro, Viale dell’Universit\`a 2, I-35020 Legnaro (Italy)}
\affil[7] {INFN - Sezione di Milano, Via Celoria 16, I-20133 Milano (Italy)}
\affil[8] {Universit\`a di Bari, Dipartimento Interateneo di Fisica ‘M. Merlin’, Via G. Amendola 173, I-70126 Bari (Italy)}
\affil[9] {INFN - Sezione di Bari, Via E. Orabona 4, I-70125, Bari (Italy)}
\affil[10] {Politecnico di Torino, Dipartimento DISAT, Corso Duca degli Abruzzi 24, I-10129 Torino}
\affil[11] {INFN - Sezione di Torino, Via P. Giuria 1, I-10125 Torino (Italy)}
\affil[12] {INFN - Sezione di Catania, Via Santa Sofia 64, I-95123 Catania, Italy}
\affil[13] {INFN - Sezione di Ferrara, Via Saragat 1, I-44122 Ferrara (Italy).}
\affil[14] {INFN - Laboratori Nazionali del Sud, Via Santa Sofia 62, I-95123 Catania (Italy)}
\affil[15] {INFN - Sezione di Perugia, Via A. Pascoli, I-06123, Perugia (Italy)}
\affil[16] {INFN - Laboratori Nazionali di Frascati, Via E. Fermi 40, I-00044 Frascati (Italy)}
\affil[17] {INFN - Sezione di Napoli, Via Cinthia, I-80126 Napoli (Italy)}
\affil[18] {Universit\`a di Napoli ``Federico II'', Dipartimento di Fisica ``E. Pancini'', Via Cinthia 21, I-80126 Napoli (Italy)}
\affil[19] {Universit\`a di Trieste, Dipartimento di Fisica, Via A. Valerio 2, I-34172 Triste (Italy)}
\affil[20] {INFN - Sezione di Trieste, Via A. Valerio 2, I-34172 Triste (Italy)}
\affil[21] {Universit\`a di Milano, Dipartimento di Fisica, Via Celoria 16, I-20133 Milano (Italy)}
\affil[22] {INFN - Sezione di Bologna, Viale C. Berti Pichat 6/2, I-40127 Bologna (Italy)}
\affil[23] {Universit\`a di Bologna, Dipartimento di Fisica e Astronomia, Viale C. Berti Pichat 6/2, I-40127 Bologna (Italy)}
\affil[24] {ENEA - Bologna, Via Martiri di Monte Sole 4, I-40129 Bologna (Italy)}
\affil[25] {Universit\`a di Padova, Dipartimento di Fisica e Astronomia, Via Marzolo 8, I-35131, Padova (Italy)}
\affil[26] {Universit\`a di Perugia, Dipartimento di Fisica e Geologia, Via A. Pascoli, I-06123, Perugia (Italy)}
\affil[27] {INFN - Sezione di Roma, p.le Aldo Moro 2, I-00185 Roma (Italy)}
\affil[28] {Universit\`a di Catania, Dipartimento di Fisica e Astronomia, Via Santa Sofia 64, I-95123 Catania, Italy}

\date{}
\maketitle
\newpage

\section*{Summary}
\vspace{-0.5cm}
Nuclear astrophysics, the union of nuclear physics and astronomy, went through an impressive expansion during the last twenty years. This could be achieved thanks to milestone improvements in astronomical observations, cross section measurements, powerful computer simulations and much refined stellar models.\\
Italian groups are giving quite important contributions to every domain of nuclear astrophysics, sometimes being the leaders of worldwide unique experiments. In this paper we will discuss the astrophysical scenarios where nuclear astrophysics plays a key role and we will provide detailed descriptions of the present and future of the experiments on nuclear astrophysics which belong to the scientific programme of INFN (the National Institute for Nuclear Physics in Italy).

\tableofcontents

\markright{Introduction}
\section{Introduction}
Different fields are intimately linked together in nuclear astrophysics. As a matter of fact, nuclear physics is shaping the Universe from the structure of the nuclei to the abundances of the chemical elements, from nuclear reactions to the life and death of stars.
In this paper we will first discuss the scenarios where nuclear astrophysics plays a fundamental role, shortly describing the activities 
which are addressing each of the different scenarios and 
which belong to the scientific program of the National Institute for Nuclear Physics in Italy (INFN) and, in particular, to Commissione Scientifica Nazionale  III for experimental researches on nuclear physics (CSN3). Then, we will provide rather detailed status and prospects  of the experiments which are essentially devoted to nuclear astrophysics and of a few other experiments which are contributing or are expected to contribute in the near future to nuclear astrophysics. Finally, the equation of state of dense and ultra-dense matter is discussed together with several experiments contributing to its study.\\
We point out that this paper is representing neither an exhaustive nor a priority list of the items which will be financed by INFN in the future. We are aware of probably not covering all the experiments which gave a single important contribution to nuclear astrophysics: we apologize with them in advance. We also do not discuss the tools of nuclear theory applied to nuclear astrophysics: it is a wide and important subject which, to our opinion, would deserve its own review.

\markright{BBN}
\section{Big Bang nucleosynthesis}
It has been early recognized that primordial nucleosynthesis provides a unique probe of the early Universe \cite{peebles1993}.
By comparing the abundances of light elements measured in pristine material, in stars or diffuse matter,
to the predictions of cosmological models,
we may test these models as well as the laws of fundamental physics which are at the heart of them.
For instance, the
primordial He abundance depends on the expansion rate of the Universe during the first 3 minutes since the ``bang''.
Therefore, a discrepancy between the measured abundances of the main products of the Big Bang nucleosynthesis (BBN) and the predictions of a
standard cosmological model may be interpreted as an evidence of new physics, such
as the existence of non-standard relativistic particles contributing to the expansion rate
of the primordial Universe (see, e.g., \cite{izotov2014}).
In addition, the primordial deuterium and lithium
are both sensitive to the baryons-to-photons density ratio ($\eta$). In this context, while the observed He
and D abundances appear in good agreement with the value of $\eta$ independently derived from the
temperature fluctuations of the
cosmic microwave background \cite{planck2014}, the Li measured in very metal poor stars of our Galaxy,
the so-called Spite plateau,
would require a substantially lower value of $\eta$. So far, the various attempts made to solve this ``Li problem''
invoking an erroneous interpretation of the observed Spite plateau were unsuccessful, so that a revision of the
standard cosmological model
is not excluded. In any case, the BBN predictions rely on accurate determinations of the relevant nuclear
reaction rates. Note that in the case of the BBN calculation, the reaction rates should be known with an uncertainty of
just a few percent, a precision  much higher than that usually required for stellar nucleosynthesis calculations.
This occurrence is a challenge for nuclear astrophysics.\\
\mbox{} \newline
A minimal nuclear network for BBN calculations includes about 8 isotopes and 12 reactions.
Two of these reactions have been studied in the underground Gran Sasso Laboratory (LNGS) by LUNA (Laboratory for Underground Nuclear Astrophysics) inside the energy region of astrophysical interest taking advantage of the background suppression: $^{3}$He($\alpha$,$\gamma$)$^{7}$Be and $^{2}$H($\alpha$,$\gamma$)$^{6}$Li. The former has also been studied by ERNA (European Recoil separator for Nuclear Astrophysics) at higher energies.
Moreover, the reactions $^{2}$H($^{2}$H,p)$^{3}$H, $^{2}$H($^{2}$H,n)$^{3}$He, $^{3}$He($^{2}$H,p)$^{4}$He and $^{7}$Li(p,$\alpha$)$^{4}$He have been studied by AsFiN with the Trojan Horse Method (THM). THM is a powerful indirect method, which requires normalization to high energy data but which allows for the exploration of the very low energy region also removing the electron screening effect.\\
Results are expected in the near future by LUNA on  $^{2}$H(p,$\gamma$)$^{3}$He and by AsFiN with the THM on $^{3}$He($\alpha$,$\gamma$)$^{7}$Be, $^3$He(n,p)$^3$H, $^7$Be(n,$\alpha$)$^4$He and $^7$Be(n,p)$^7$Li. \\
Another series of key experiments for BBN studies has been carried out at n$\_$TOF, the neutron Time-of-Flight facility at CERN, where the $^7$Be(n,p)$^7$Li and $^7$Be(n,$\alpha$)$^4$He reactions have been directly measured for the first time in a wide energy range, providing additional constrains for the reaction rates used in BBN. The published results of all these experimental efforts have excluded the nuclear solution to the controversial cosmological $^6$Li problem
and have ruled out the neutron-induced reactions as potential explanation of the cosmological $^7$Li problem. Furthermore, $^{2}$H($^{2}$H,p)$^{3}$H and $^{2}$H($^{2}$H,n)$^{3}$He are presently studied after cluster Coulomb explosion in laser-induced plasma by AsFiN.\\
Finally, anti-nuclei production is studied by ALICE at CERN LHC and it will be refined in the future.

\markright{Hydrostatic Burning Phase}
\section{Nucleosynthesis in hydrostatic burnings}

Stars are self-gravitating objects in thermal and hydrostatic equilibrium during most of their lifetime.
They evolve because they lose energy, from the surface (radiation)
or from their hot interiors (thermal neutrinos). Then,
to maintain the hydrostatic and thermal equilibrium, the stellar core must contract and,
so doing, extracts energy from its gravitational field. However, the amount of
gravitational energy released in this way exceeds the energy leakage,
so that the internal (thermal) energy must increase.
Then, when the temperature becomes large enough,
thermonuclear reactions enter into the game. Summarizing, stellar evolution is controlled by an
interplay between gravity and nuclear reactions leading to a succession of short
phases, during which the stellar core contracts and heats up, and longer phases, during
which nuclear reactions are in action.

\subsection{The H-burning phase}
The first hydrostatic burning phase
proceeds in different ways depending on the stellar mass. Low-mass stars, like the
Sun, burn hydrogen through the pp chain, while in intermediate-mass and massive stars the
H burning proceeds through the CNO cycle. In the former case, the leading reaction is the $p+p\rightarrow D+e^++\nu$,
which controls the H-burning rate. Its cross section is too low to be measured in laboratory conditions,
so that theoretical estimations are currently used in stellar models calculations. However, measurements are
possible for other reactions of the pp chain, which are of primary interest to understand the solar-neutrino flux at Earth and
for the nucleosynthesis of some light isotopes, such as $^3$He and $^7$Li (see, e.g., \cite{solarfusionII}). Concerning the CNO cycle,
the leading reaction is $^{14}$N(p,$\gamma$)$^{15}$O, which has been studied at LUNA. Other reactions
involving C, N and O isotopes are of great interest for understanding the isotopic composition observed in
giant stars of different mass.
Indeed, when the CNO reaches the equilibrium, as it happens in the
H-burning shell of a giant star, an isotopic ratio is simply given by the ratio of the respective production/destruction rates.
An example is the abundance ratio of $^{16}$O and $^{17}$O.
In practice, since the reaction rates only depend on temperature,
this isotopic ratio traces the temperature profile within the H-burning zone.

When the ashes of the H-burning are mixed to the stellar surface due, for
instance, to convective instabilities affecting the stellar envelope in red
giant (RGB) and asymptotic giant (AGB) stars, variations of these isotopic composition appear at
the stellar surface. This isotopic ratios can be routinely measured from
IR spectra of giant stars \cite{abia2012,lebzelter2015,hinkle2016}.
These measurements allow us
to probe the depth attained by the mixing episodes responsible for the
chemical variations observed at the stellar surface. We recall that in addition
to convection, other processes may induce  or contribute to deep mixing
episodes, such as rotation, magnetic buoyancy, gravity waves, thermohaline circulation \cite{dearborn1992, charbonnel1995, denissenkov2003, palmerini2011}. Therefore, we can learn a lot about these
phenomena by looking at the isotopic composition of giant stars.

\subsection{Ne-Na and Mg-Al cycles}
In the hottest layers of a H-burning zone (T $> 30-40$ MK), other two cycles may be active. The first involves isotopes of Ne and Na,
the second those of Mg and Al. Globular Clusters are building blocks of galactic halos. They are found in any kind of
galaxy, spirals or ellipticals, as well as in dwarfs or irregular galaxies.
The well known anti-correlation between O and Na observed in the majority of the
galactic Globular Clusters is an evidence of primordial pollution from massive or
intermediate-mass stars where, in addition to the O depletion due to the CNO cycle, a sodium production by
the Ne-Na cycle was active (see, e.g., \cite{gratton2012}). An anti-correlation between Mg and Al is
also found. This chemical pattern is a fossil record of the formation epoch of these stellar systems
and may be used to understand the complex phenomena leading to galaxy formation.
A further reason for studying
these cycles is the possibility to produce radioactive isotopes with a relatively long half-life.
For example, the presence of radioactive $^{26}$Al
(ground state half-life $t_{1/2}\approx7\times10^5$ yr)
in different astronomical environments may be a trace
of the operation of the Mg-Al cycle in
stellar interiors. The detection of the 1.809 MeV $\gamma$-ray line, emitted by the decay of the
excited daughter $^{26}$Mg, demonstrates
that a few M$_\odot$ of $^{26}$Al is presently alive in the galactic disk
\cite{diehl2006}. The Ne-Na and Mg-Al cycles are active in the H-burning core of massive stars.
These stars undergo intense mass loss episodes and may end their life as core-collapse supernovae.
Therefore, they provide an important contribution to the pollution of the interstellar medium.
For this reason, the amount of $^{26}$Al in the galactic disk constrains this pollution and, in turn,
the star formation rate.
In addition,
the excess of $^{26}$Mg found in solar system material,
proves that some radioactive $^{26}$Al, perhaps contained in the ejecta of a nearby supernova,
has been injected into the presolar nebula shortly
before the solar system formation, about 4.5 Gyr ago \cite{lee77}.

\subsection{Advanced burning}
The final fate of a star essentially depends on the mass of its C-O core \cite{arnett}.
For example, it exists a critical core mass for the C ignition. Stars that develop
a core more massive than this critical value, ignite C and terminate their life with a core collapse,
eventually followed by a supernova explosion. In any case the remnant may be either a neutron star or a black hole.
On the contrary less massive stars skip the carbon ignition and enter the
AGB phase, during which they lose the whole envelope, leaving a bare CO white dwarf (WD).
This critical core mass, however, depends on the C/O ratio within the core and on the C-burning reaction rate.
The first quantity, the C/O ratio left in the core at the end of the He-burning, is the result of the
competition between the C production, as due to the
triple-$\alpha$ process, and its destruction, as due to the $^{12}$C($\alpha,\gamma$)$^{16}$O reaction.
While the rate of the triple-$\alpha$ is known within a 10\% error, in the range of He burning temperatures \cite{fynbo}, the
$^{12}$C($\alpha,\gamma$)$^{16}$O reaction rate is unknown within a factor of 2 \cite{circe2013}.
As a consequence, the  uncertainty of the resulting C/O ratio is rather large (see, e.g., \cite{imbriani2001}).\\
Concerning the C burning rate, the leading process is  the $^{12}$C$+^{12}$C reaction, whose cross section deserves more studies for center of mass energies $E_{CM} < 2$ MeV.
An improved determination of the critical CO core mass is mandatory for a large variety of astrophysical issues. For instance,
the rates of the different classes of supernovae depend on this mass boundary. Above this limit there are the progenitors of
core-collapse supernovae (type IIL, IIN, IIP, Ib, Ic and e-capture supernovae), while below this limit
there are the progenitors of thermonuclear supernovae (type Ia).

\subsection{Beyond iron: Nucleosynthesis of heavy nuclei}
\begin{figure}[ht]
\resizebox{1\textwidth}{!}{\includegraphics{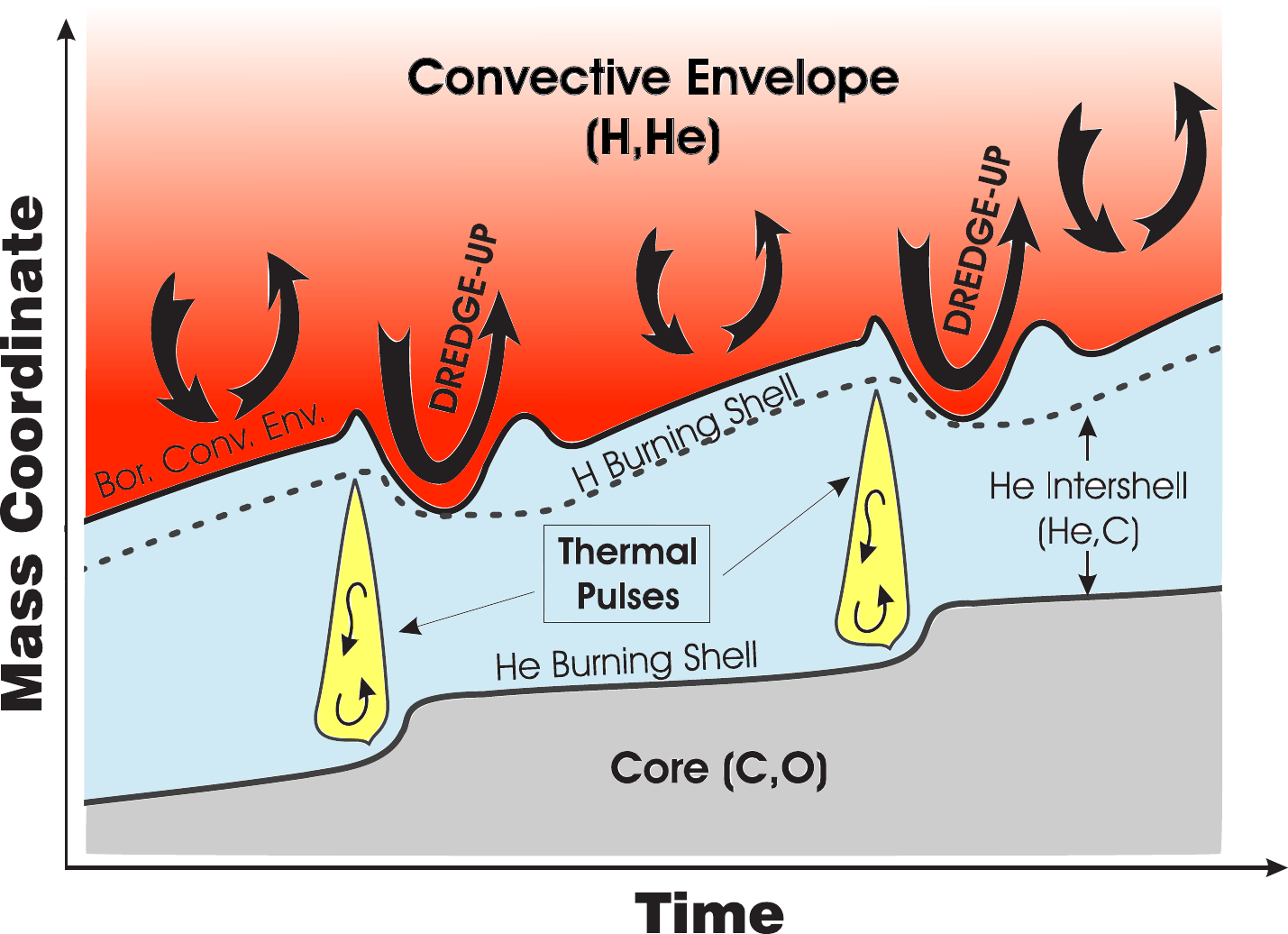}}
\caption{This sketch illustrates how the positions of the inner border of the convective envelope, the H-burning shell and the He-burning shell evolve during the thermally pulsing AGB phase. The convective regions generated by two subsequent thermal pulses are also shown. The s-process nucleosynthesis occurs between two pulses within a thin region enriched in $^{13}$C and located just below the layer of maximum penetration of the convective envelope.}
\label{cartoon}
\end{figure}

With the exception of a bunch of rare isotopes, all the elements beyond the iron group are produced
through neutron captures. A series of slow neutron-capture episodes (s-process) takes place in the He-rich mantle of
AGB stars undergoing recursive thermal pulses (see Fig. \ref{cartoon}, for a review see \cite{straniero2006}).

This process accounts for most of the stable isotopes with magic neutron numbers (50, 82 and 126) and mass number A greater than 90, such as $^{90}$Zr, $^{138}$Ba, $^{139}$La, $^{142}$Nd, $^{208}$Pb.
The main source of neutrons is the $^{13}$C($\alpha$,n)$^{16}$O reaction, which sustains a particularly low neutron density
($\sim 10^7$ neutrons per cm$^3$) for a rather long period (about $10^5$ yr). These conditions guarantee a large
enough neutron exposure to allow the production of the heaviest s-process nucleus ($^{209}$Bi).
The lighter s-process elements (up to the $^{88}$Sr bottleneck) are instead produced by massive stars, during the late part of the core-He burning or in their C-burning shells.
In that case the neutron source is the $^{22}$Ne($\alpha$,n)$^{25}$Mg.

It is important to stress that the s-process synthesizes only half of the
elements heavier than iron. The remaining isotopes are created by the rapid
neutron capture process (the r-process, see Section 4.3), which is
characterized by extremely large neutron densities ($n_n>10^{23}$
cm$^{-3}$). However, the theoretical knowledge of the r-process is
not evolved as its s-process counterpart. As a consequence, the r-process
contribution to the solar inventory is commonly calculated as a residual
from its slow neutron capture process counterpart. Therefore, very precise neutron capture cross sections for heavy
isotopes are needed to exclude any possible source of uncertainty. This is
particular important for the study of s-only isotopes since
they allow to test the robustness of stellar models.

\mbox{} \newline
In the last twenty years the Italian astro-nuclear community, in particular LUNA and AsFiN, has put a major effort in
improving our knowledge of almost all the key nuclear processes of the proton-proton chain and of the CNO cycles, complemented by ERNA at higher energies.
In particular, LUNA has obtained
the first direct determination of the cross section of $^{3}$He($^{3}$He,2p)$^{4}$He within the Solar Gamow peak definitively ruling out the nuclear solution to the solar neutrino problem. Eventually, fundamental inputs have been provided to the standard solar model for an accurate prediction of the solar neutrino spectrum.
The results from LUNA on the bottleneck reaction of the CNO cycle, $^{14}$N(p,$\gamma$)$^{15}$O, reduced by a factor of two the CNO neutrino flux expected from the Sun and increased by about 1 Gyr the limit on the age of the Universe obtained through the analysis of the Globular Cluster turnoff point.\\
The studies of selected key reactions of the cycles responsible for hydrogen burning at temperature higher than the one of our Sun, CNO, Ne-Na and Mg-Al, have squeezed the error on the prediction of the light isotope abundances and have allowed
for the firm identification of the production site of a few key isotopes. The remaining few important reactions of the hydrogen burning are going to be studied in the near future.\\
Both the fundamental reactions of the advanced burning, $^{12}$C($\alpha,\gamma$)$^{16}$O and $^{12}$C$+^{12}$C, are addressed by INFN groups.
The former is being studied by ERNA and it will be studied underground by LUNA MV with a new 3.5 MV accelerator, the latter has been studied both by ERNA and, in the very low energy region, by AsFiN with the THM. In particular, the $^{12}$C$+^{12}$C cross section has been indirectly measured by AsFiN with an uncertainty of about 30$\%$ and found to experience a strong resonant behavior in the region down to 1 MeV. This energy region will be explored in the near future by LUNA with a direct measurement at the 3.5 MV underground accelerator.\\
By using CHIMERA (Charged Heavy Ions Mass and Energy Resolving Array) and FARCOS (Femtoscope ARray for COrrelations and Spectroscopy), studies have been performed of pygmy resonances \cite{cardella2017}  and of branching ratio for $\gamma$ decay  \cite{martorana2018} of $^{12}C$ states involved in the helium burning phase and in general in the creation of carbon in the Universe. Studies on excited levels rare decays of medium light ions, important for astrophysical environments, will be pursued.
With stable beams available at Italian laboratories, INFN researchers performed experiments aimed to better determine the $S$-factor of the $^{10}$B(p,$\alpha$)$^{7}$Be \cite{lombardo_jpg} and $^{19}$F(p,$\alpha$)$^{16}$O reactions \cite{lombardo_bull, He-2018}. In particular, a new estimate of the $^{19}$F(p,$\alpha$)$^{16}$O reaction rate at AGB temperatures has been obtained \cite{lombardo_plb}, contributing to a better understanding of the fluorine nucleosynthesis in stars. In another experiment, the structure and decay pattern of the Hoyle state, that is involved in $^{12}C$ nucleosynthesis via the triple-alpha process \cite{nguyen}, was carefully studied at LNS \cite{dellaquila_prl} with a solid state hodoscope derived from the OSCAR (hOdoscope of Silicons for Correlations and Analysis of Reactions) project \cite{dellaquila_nim}. \\ 
Finally, the nuclear ingredients of the s-process are addressed from several different view points. $^{13}$C($\alpha$,n)$^{16}$O, one of the two most important neutron sources inside stars,  has been studied by AsFiN down to very low energies with the THM and it is being measured underground in the low energy region by LUNA at the 400 kV accelerator (to be followed by a measurement over a wider energy region at the 3.5 MV accelerator). A complementary approach to the above described methods will come from the n$\_$TOF experiment, which will provide information on the $^{13}$C($\alpha$,n) reaction by studying the inverse reaction $^{16}$O(n,$\alpha$)$^{13}$C and focusing on specific $^{17}$O states.
The same is true for the other neutron source, $^{22}$Ne($\alpha$,n)$^{25}$Mg: within INFN the first hints on this cross sections came from high-resolution measurements by the n$\_$TOF experiment on the $^{25}$Mg(n,$\gamma$) and $^{25}$Mg(n,tot) cross sections, which allowed an unambiguous spin/parity assignment of the corresponding excited states in $^{26}$Mg. Data have also been taken by AsFiN, and in the next years additional information will be provided by LUNA MV.\\
Among the fundamental inputs for the s-process nucleosynthesis calculations there are the neutron capture cross sections, especially on branching-point nuclei. A large effort on this item has been done and it will be carried out in the future by the n$\_$TOF collaboration with direct measurements, whereas GAMMA and SPES (Selective Production of Exotic Species, at Legnaro National Laboratories) will join with indirect measurements. The most important poison reactions for the s-process, $^{14}$N(n,p)$^{14}$C and $^{26}$Al(n,p)$^{26}$Mg, have already been measured at n$\_$TOF (data are currently under analysis) and can be measured indirectly by AsFiN. Finally, contributions are expected to come from PANDORA (Plasmas for Astrophysics Nuclear Decays Observation and Radiation for Archaeometry), with the measurement of the half-life of
crucial branching point nuclei in plasma.

\markright{Explosive nucleosynthesis}
\section{Explosive Nucleosynthesis}

\subsection{Novae and X-ray bursters}
Novae occur in close binary systems made by a white
dwarf and a low-mass main-sequence star \cite{boev2008}. Due to their close proximity,
the main sequence star overflows its Roche lobe (the region around a star in a binary system within which orbiting material is gravitationally bound to that star) and hydrogen is slowly accreted
onto the WD surface.
Owing to the slow accretion, the H remains cold, allowing the development
of a mild electron degeneracy. When a critical H mass is attained,
a thermonuclear runaway occurs.  The following electromagnetic burst is characterized
by  a  sudden  increase  of  the  brightness  by $\sim 10$  magnitudes
in  a  few  days.
\begin{figure}[ht]
\includegraphics[width=12cm,trim={0cm 0cm 4cm 3cm},clip]{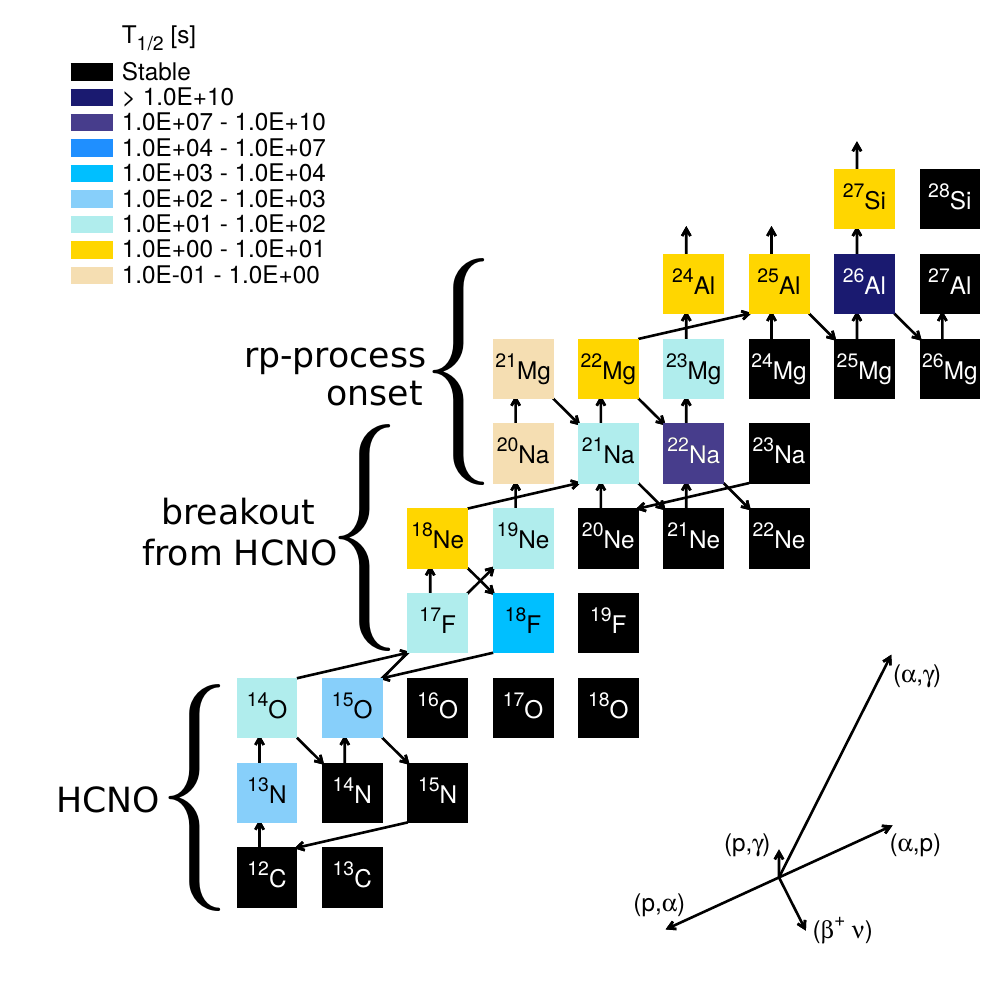}
\caption{Nuclear network of a hot CNO burning (HCNO). The CNO breakout and the rp-process are active at the higher temperatures only.}
\label{hotcno}
\end{figure}
The  typical  maximum  luminosity  is  about
$10^4-10^5$ L$_\odot$. The decline in luminosity, back to the original
level,  takes  several  days  to  a  few  months.  Following  the
explosion,  most  of  the  accreted  material  is  ejected  into  the
interstellar medium. In such a
way the nucleosynthesis induced by this phenomenon
provides a significant contribution to the chemical evolution of Galaxies.
The maximum temperature attained during the burst depends on the WD mass: it ranges
between 200 and 400 MK.
In any case, the temperature is high enough to allow a fast (or hot) H burning.
At variance with
the hydrostatic H burning, proton captures on radioactive nuclei
compete with $\beta$ decays. Thus, nuclear chains usually
forbidden in hydrostatic burning are here activated (see Fig. \ref{hotcno}).
For example, the hot
CN cycle may proceed trough
$^{12}$C(p,$\gamma$)$^{13}$N(p,$\gamma$)$^{14}$O$\rightarrow ^{14}$N,
thus by-passing the $^{13}$C production that occurs when the usual chain,
$^{12}$C(p,$\gamma$)$^{13}$N$\rightarrow ^{13}$C(p,$\gamma$)$^{14}$N, is followed.
At the high temperature of a nova explosion, important branch points are open,
such as the $^{17}$O(p,$\gamma$)$^{18}$F, that becomes competitive with the
$^{17}$O(p,$\alpha$)$^{14}$N and allows the production of the radioactive $^{18}$F. \\
Recurrent thermonuclear bursts are also observed in low-mass X-ray binaries.
In this case the compact component is a neutron star.
As H is accreted on the neutron star surface, it is immediately converted into He.
Later on, when the critical He mass is attained, an explosive He burning occurs.
In this case the peak temperature may be T $> 500$ MK, so that rapid proton
captures (rp-process) extend
to heavier nuclei (Al, Si, Sc).

\subsection{Supernovae}
Two different explosion mechanisms are known for the observed supernovae (SNe).
In type Ia supernovae, the engine is the C ignition in the dense core of a CO WD.
After the C ignition, a rather long phase of
quiescent C burning takes place (for a few kyr). Then, when the central density approaches
$\sim 10^{6}$ g/cm$^{-3}$, a flame, starting from the center, propagates outward on a dynamical timescale.
At the beginning, the flame speed is sub-sonic (deflagration), but at a certain point it should
become super-sonic (detonation) \cite{AK91}.
The most common channels through the
type Ia explosion  are the so-called single-degenerate, due to
mass transfer in close binary systems made by a CO WD and a main-sequence or giant companion,
and the double-degenerate, as due to WD+WD mergers.
In both cases the C ignition occurs when the primary WD exceeds the Chandrasekhar
mass limit. Another known channel for type Ia SNe is the so called sub-Chandrasekhar.
 It requires a binary system made by a CO WD accreting He from an He-rich star companion. When a critical He mass is transferred,
a violent He detonation occurs, causing the propagation of a shock wave toward
the center. Then, when the shock arrives at the center, a C detonation occurs.
The most important nuclear physics inputs for understanding type Ia explosions are the
rates of the leading processes: the $^{12}$C$+^{12}$C reaction, electron captures and $\beta$ decays.
In particular, the weak interactions determine the degree of neutronization
of the nuclear matter,
which is a fundamental parameter affecting the physical conditions
(temperature and density) at the onset of the explosion (dynamical breakout).\\
The second explosion mechanism is the collapse of the degenerate core of a massive star.
Also in this case the weak interactions play a fundamental role.
Indeed, in case of strong degeneracy, electron captures are favoured with respect
to $\beta$-decays.
The consequences of the electron captures are dramatic: the pressure by degenerate electrons
suddenly drops and the core collapses.
Once the core density approaches the nuclear density ($\sim 10^{14}$ g/cm$^3$),
matter in the core is almost completely converted into degenerate neutrons, whose pressure is large enough to suddenly
stop the collapse. As a result of the weak interactions, about $10^{53}$ erg of gravitational energy
are converted into neutrinos.
Then, a stable core of about 0.8 M$_\odot$ forms, sustained by the pressure of degenerate neutrons.
Meanwhile, the still falling external layers bounce on the hard surface of the proto-neutron star giving rise, in some cases,
to a supernova event.
Hydrodynamic models of this process (see \cite{WJ2005} and references therein)
show that a shock wave is initiated which propagates outward.
However, as the shock front moves outward, the temperature suddenly rises, so that the in-falling material passing
through the shock photo-dissociates.
As a consequence, the shock wave loses energy and stalls. Nevertheless, the partial deposition of the neutrinos energy
may revitalize the shock. Thus, streaming out neutrinos from the inner neutron star are needed to
enable a supernova explosion.\\
As the shock propagates outward the local temperature rises thus inducing direct and reverse nuclear reactions.
Most of this material is ejected at high speed in the interstellar medium.
Therefore, supernovae are the major contributors to the chemical evolution. For instance about 2/3 of
the iron in the solar system has been produced by type Ia supernovae, and 1/3 by core-collapse supernovae.
The core collapse is also responsible for the production of a compact remnant, neutron star or black hole.

\subsection{r-process in neutron star mergers}
The heavy isotopes not produced by the s-process are instead produced through rapid neutron captures (r-process).
The first evidence of heavy elements produced via rapid neutron captures was obtained by  analyzing data of an
USA H-bomb test on the Bikini Atoll in 1950.
\begin{figure}[htb]
\begin{center}
\resizebox{0.8\textwidth}{!}{\includegraphics{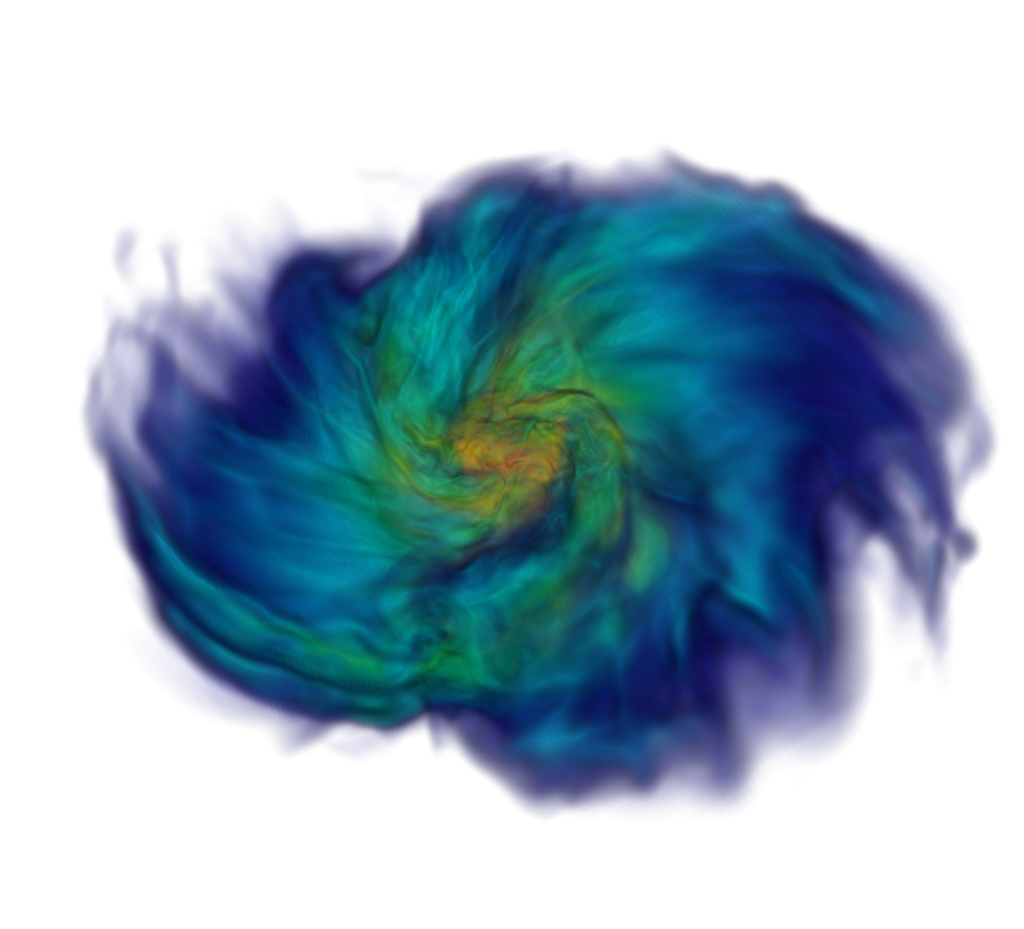}}
\caption{3D density rendering of a post-merger system (two neutron stars with M=1.35 M$_\odot$), extracted from a relativistic simulation including a real equation of state and neutrino transport (courtesy of D. Radice; see \cite{Radice-2018} for details)}
\label{nsmerger}
\end{center}
\end{figure}
The connection with stellar explosions immediately became obvious \cite{B2HF}.
In the last few years a growing amount of evidence indicates that the bulk of the r-process nucleosynthesis occurs during
the merging of small compact objects, such as two binary neutron stars or a neutron star and a black hole
(see Fig. \ref{nsmerger}).
In particular,
the kilonova (or macronova) phenomenon observed after a few days since the occurrence of a short-gamma-ray burst has
been interpreted as due to the decay of a few $10^{-3}$ M$_\odot$ of heavy radioactive nuclei
produced by the r-process nucleosynthesis (for a review see \cite{tielemann2016}) taking place within the dynamical ejecta of a
binary neutron star merger. This scenario has been reinforced by the recent detection by the LIGO and Virgo Collaborations of a
gravitational wave signal from a BNS merger (GW170817), whose electromagnetic counterpart was
identified as a short-gamma-ray burst (GRB170817A) by the LAT instrument on board of the NASA's FERMI satellite, followed, after about 9 days, by the
optical and IR observation of an associated  kilonova (SSS17a).
At variance with the s-process,
which always proceeds close to the $\beta-$stability valley, r-process
involves nuclei far from the stability and close to the neutron drip line. In particular, it implies higher neutron densities, $>10^{20}$ neutrons per cm$^3$ (see e.g. \cite{Perego-2017}). Rather than the neutron capture cross sections, other nuclear data play a fundamental role
in the r-process calculations, among which
masses of nuclei far from stability, properties of $\beta$ decay and electron/positron captures on
nuclei and nucleons, fission barriers and fission fragment distributions.\\
\mbox{} \newline
Most of the reactions taking part in the explosive burning of novae and X-ray bursters are also involved, at lower energies, in the hydrostatic burning and are studied by the same groups (see sec. 3.1 and 3.2). The same is true for $^{12}$C$+^{12}$C, the fundamental reaction responsible for the explosion of thermonuclear supernovae (SNe Ia, the standard candles of Cosmology). Here the energy region of interest in the center of mass extends down to 0.7 MeV, it has been covered by AsFiN with the THM down to 1 MeV and it will be studied in the future by LUNA with the 3.5 MV underground accelerator down to about 1.6 MeV and by ERNA down to 2.0 MeV.\\
The abundance of the r-process elements in a star is obtained from the difference between the measured abundance of a given element and the abundance expected from the s-process. As a consequence, all the efforts previously summarized to improve the knowledge of the s-process are also contributing
to better understand the r-process.
It is worth to note the important role played on this subject by the n$\_$TOF experiment, which provides high-accuracy data on neutron-induced fission cross sections of a series of transuranic isotopes (U, Pu, Th, Am, Cm, Cf). Such data are fundamental to derive  physical properties of neutron-rich isotopes involved in the fission recycling process during neutron stars mergers. In particular, the shape of the third r-process peak strongly depends on these inputs.
In addition to this, important contributions are expected to come in the near future from GAMMA and SPES through the study of shapes/deformation and resonance (pygmy) excitation modes of exotic species lying along or in the vicinity of the r-process path and
neutron capture cross sections through indirect methods, respectively.

\markright{Nuclear Experiments}
\section{Nuclear experiments}

\subsection{AsFiN}

The AsFiN group has been active in the field of nuclear astrophysics since
the early 90s, mostly using indirect methods to measure nuclear reaction
cross sections of astrophysical interest. The focus on indirect methods
has also implied extensive investigations of reaction mechanisms, whose
understanding is pivotal in relation to the application of nuclear reaction
models to the extraction of the cross sections of astrophysical importance.
Such investigations have been carried out both theoretically and experimentally,
in collaboration with scientists from over fifteen countries.

\mbox{} \newline
The main line of research of the AsFiN collaboration is the investigation of
reactions of astrophysical relevance using the Trojan Horse Method (THM) \cite{SPITALERI91}. In
this framework, the cross section of a reaction with three particles in the
exit channel is measured to deduce the cross section and, consequently, the
astrophysical factor, of a reaction of astrophysical interest. In detail, a
$A(x,b)B$ reaction where the involved nuclei are charged nuclei or neutrons,
can be investigated by using the $A(a,bB)s$ process where particle $a$ shows
a $x+s$ cluster structure. Under peculiar kinematic conditions, $a$ breakup
takes place in the nuclear field of $A$ and while the participant $x$ induces
the nuclear reaction of astrophysical interest, $s$ is emitted without
influencing the $A(x,b)B$ sub reaction. Since $a$ breakup takes place inside
$A$ nuclear field, the $A-x$ interaction is not suppressed by the Coulomb
barrier penetration, which has been already overcome in the entrance channel,
making it possible to reach zero energy with high precision. It takes its roots on exended studies of quasi-free processes dating from the early eighties and a comprehensive view is given in \cite{TRI14}.\\
Similar considerations apply to the Asymptotic Normalization Coefficient technique (ANC), which has been shown to have a tight
connection with THM \cite{LAC12}. In this case, the transfer to a bound state
is studied to extract the ANC and calculate the cross section of radiative capture
reactions at energies well below the Coulomb barrier \cite{TRI14}.\\
The use of indirect methods makes it possible to perform high accuracy measurements
with generally simple experimental setups, thanks to the large energies and yields
of the indirect process to be investigated. In the case of reactions induced by
radioactive nuclei, the problem of the vanishing signal-to-noise ratio is even more
severe owing to the low beam intensities currently available. In this case, also the
application of indirect methods calls for more complicated, large-surface detection
arrays such as TECSA \cite{PIZ16} and ASTRHO \cite{CHE15}. In particular, the Thick-Target
Inverse Kinematics (TTIK) approach \cite{ART90} has turned out very useful to measure a
whole excitation function of an elastic scattering process especially involving
radioactive nuclei, allowing one to perform nuclear spectroscopy of states involved
in peculiar astrophysical scenarios (see, for instance, \cite{TOR17}).\\
Finally, the availability of high-power lasers and of high-resolution, high-luminosity
$\gamma$-beams will open new opportunities in nuclear astrophysics. First, petawatt
lasers allow to induce nuclear reactions in hot plasmas, similar to the ones in
astrophysical environments. Therefore, it is possible to study \cite{QUE18} nuclear
reactions under astrophysical conditions, accounting for, e.g., electron screening
effects. Moreover, thanks to the large number of interacting ions, it would be possible
to perform direct measurements of reaction cross sections down to the energies where such
reactions take place in stars. The availability of high-intensity lasers has also made
it possible to produce high-resolution $\gamma$-beams of intensities larger than $10^6\,\gamma$/s
by means of Compton back-scattering of laser light off an accelerated counter-propagating
fast electron beam. Therefore, photodissociation reactions occurring in the presupernova
stage and in the p-process will be studied, as well as radiative capture reactions through
the detailed balance principle (for ground-state to ground-state transitions). A $4\pi$
silicon detector array is under development in collaboration with the Extreme Light Infrastructure-Nuclear Physics (ELI-NP) European Research Centre \cite{LAC17}.

\mbox{} \newline
The AsFiN group is currently active in several astrophysical context such as the primordial
nucleosynthesis, hydrostatic burning phase nucleosynthesis, s-process, classical novae.
Here is a comprehensive scheme of the present activities:
\begin{itemize}
\item The astrophysical S(E)-factor for the $^3$He($\alpha$,$\gamma$)$^7$Be reaction, which
is important for BBN and the H-burning in stellar interiors \cite{solarfusionII}
was investigated by means of the ANC technique. The experiment was performed in
Florida State University in late 2017 and is presently under analysis. Data will be
extracted in the region of astrophysical interest, as shown by the simulations. Once the
reaction rate is calculated from the present data, after comparison with other measurements
(direct and indirect) the astrophysical implications will be evaluated.
\item The study of the $^3$He(n,p)$^3$H reaction measured via THM in Notre Dame University
in 2018 is also relevant for the primordial nucleosynthesis \cite{Pizzone-2014}. The possibility, given by the THM,
to investigate neutron induced reactions was already tested in the past years \cite{gulino10},
and will allow to retrieve information on the S(E)-factor in the Gamow energy range
\item In the BBN framework AsFiN has performed two experiments in recent years to investigate
the  $^7$Be(n,p)$^7$Li and the $^7$Be(n,$\alpha$)$^4$He processes. Experiments were performed
at INFN-Legnaro National Laboratories (LNL) and at CNS-RIKEN (Japan) where the radioactive ion beam (RIB) was produced. Data analysis is
still in process and THM offers the great possibility to measure cross section of RIB interaction
with neutrons. Preliminary results were published in \cite{LAMIA17}.
\item The $^{23}$Na(p,$\alpha$)$^{20}$Ne reaction is one of the most interesting processes
in astrophysics since it is of primary importance for sodium destruction inside stars and
for the nucleosynthetic path in the A$\geq$20 mass region. This reaction is also involved
in the branching point of the Ne-Na cycle, responsible of steady hydrogen burning at high
temperatures, and is especially important for the understanding of the O-Na anti-correlations
in massive stars. The measurement has been recently carried out at the tandem accelerator of INFN-Laboratori Nazionali del Sud (LNS) using the first
$^{23}$Na beam in Europe and data analysis is ongoing.
\item The $^{27}$Al(p,$\alpha$)$^{24}$Mg reaction drives the destruction of $^{27}$Al,
the production of $^{24}$Mg and closes the Mg-Al cycle when its rate exceeds that of the
competitor reaction $^{27}$Al(p,$\gamma$)$^{28}$Si. The experiment, performed by using the
THM was completed in March 2018 at LNS tandem. Data analysis has just started.
\item Low-energy fusion reactions involving $^{12}$C and $^{16}$O are of great astrophysical
importance for our understanding of the timescale and the nucleosynthesis during late stellar
evolution of massive stars (M~$\geq$ 8 solar masses). Fusion processes like $^{12}$C+$^{12}$C,
$^{12}$C+$^{16}$O, and $^{16}$O+$^{16}$O depend not only on the reaction rate but also on
the branching between proton, neutron, or $\alpha$ decay channels. They play also a pivotal
role in the ignition of type Ia supernova and of explosive burning processes in the atmospheres
of accreting neutron stars. This is a very interesting research topic as it is demonstrated
by the Letter published in Nature by AsFiN on the pioneering study of the $^{12}$C+$^{12}$C
fusion reaction (see Fig. \ref{asfin}) \cite{TUM18}.

\begin{figure}[ht]
\resizebox{1\textwidth}{!}{\includegraphics{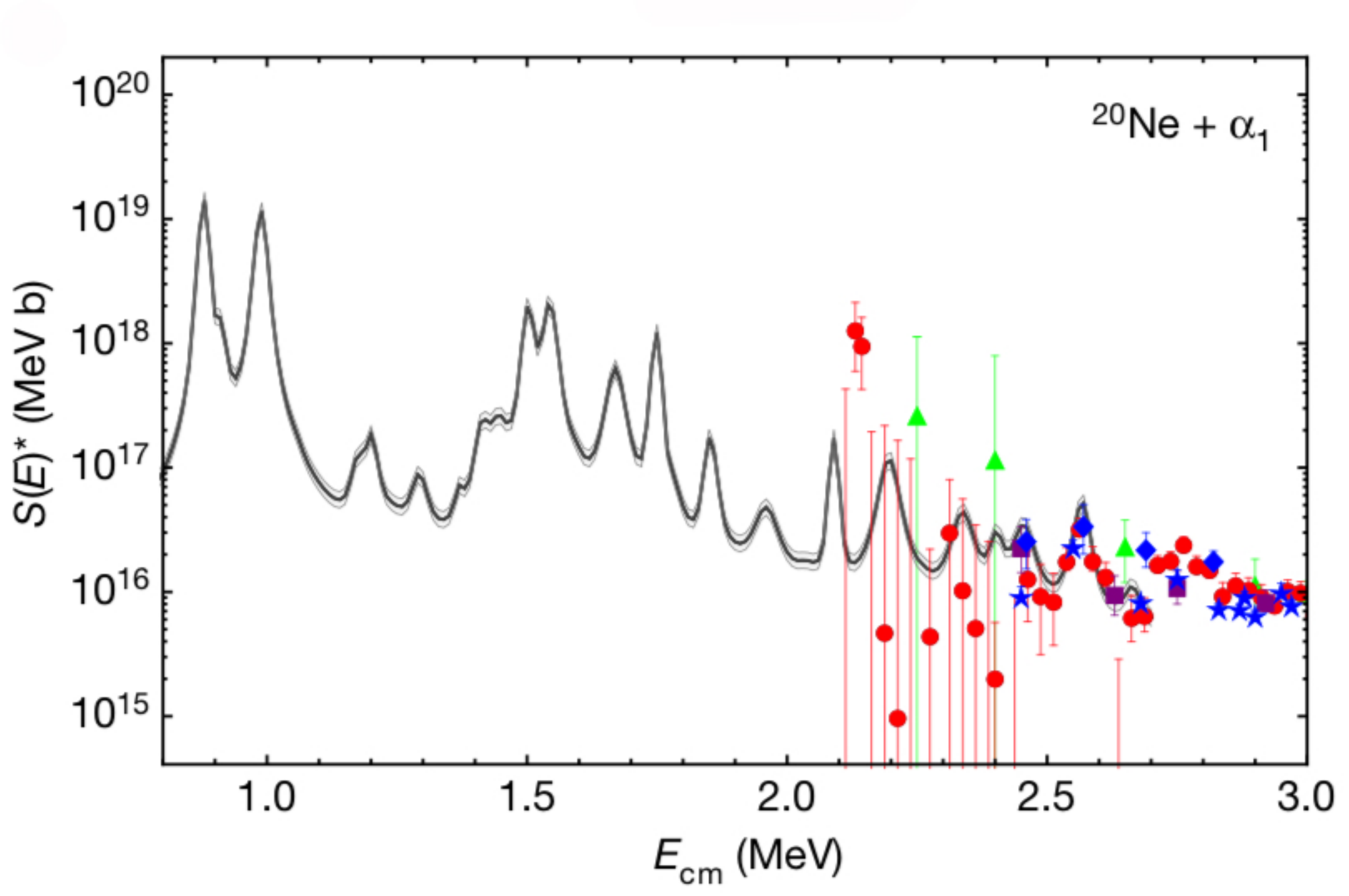}}
\caption{ The $^{12}$C+$^{12}$C astrophysical S(E)* factor for the $^{20}$Ne+$\alpha_{1}$ channel. The upper and lower grey lines mark the range arising from $\pm$1$\sigma$
uncertainties on resonance parameters plus the normalization to direct data.
Direct data are reported as
symbols (\cite{TUM18} for details.)}
\label{asfin}
\end{figure}

\item In the framework of the AGB nucleosynthesis investigation and in particular regarding
the destruction of $^{19}$F in cosmic environments, the $^{19}$F(p,$\alpha$)$^{16}$O \cite{Indelicato-2017} and
$^{19}$F($\alpha$,p)$^{22}$Ne \cite{Pizzone-2017} have been studied via the THM. The results shed light, for
the first time, on poorly known resonances at astrophysical energies. Moreover, after successfully
completing the investigation of the $^{13}$C($\alpha$,n)$^{16}$O reaction at astrophysical energies \cite{LaCognata-2013}, by supplying a robust reaction rate accounting for direct and indirect measurements, AsFiN has focused on
the $^{22}$Ne($\alpha$,n)$^{25}$Mg reaction (measurement completed in March 2018) which produces
neutrons for massive stars. Also the $^{14}$N(n,p)$^{14}$C reaction, which is the principal neutron poison in AGB stars, is presently being studied.
\item Explosive nucleosynthesis in classical novae is under investigation by the AsFiN group.
In particular the $^{18}$F(p,$\alpha$)$^{15}$O reaction was studied in recent years \cite{Cherubini-2015} and needs
further investigations especially for measuring the angular distributions. $^{18}$F decays in
the moment when the nova envelope becomes transparent to radiation, thus its $\gamma$-emission
and its abundance may probe the structure and nucleosynthesis in details \cite{LaCognata-2017}.
\item Together with the measurements performed via indirect methods reactions of astrophysical
interest are also induced in laser-driven plasma. The D(d,p)t and D(d,n)$^3$He previously investigated via the THM \cite{Tumino-2014} are presently
studied after cluster Coulomb explosion in laser-induced plasma. Data analysis is still in
progress. This measurement is of great importance as for the first time it should be possible to deduce the electron screening potential by comparing the bare nucleus S-factor measured with the THM and the screened one measured in a high temperature plasma.
\item Photodissociation plays a crucial role in many astrophysical environments such as supernovae,
massive stars, BBN and p-processes. For a prospective use of the $\gamma$-beam in ELI-NP \cite{Filipescu-2015} a first
experiment, investigating $^7$Li($\gamma$,$\alpha$)$^3$H, has been recently performed at the High Intensity Gamma-Ray Source (HIGS, TUNL) in the astrophysical energy range.
\end{itemize}

\mbox{} \newline
The AsFiN collaboration will pursue, on one hand, the investigation of nuclear reactions
of astrophysical interest using indirect methods. On the other hand, the commitment
to find novel solutions for accessing astrophysical conditions in the laboratory will be kept.
\begin{itemize}
\item Using the THM and the stable beams available at LNS including, in the next years, noble gases and pulsed beams, AsFiN will study the
reactions influencing the evolution and nucleosynthesis of massive stars, precursors of
supernovae and neutron stars, which are the subject of investigation of multimessenger
astronomy. Following the successful work on the $^{12}$C+$^{12}$C fusion \cite{TUM18}, AsFiN aims at measuring the $^{12}$C+$^{16}$O and $^{16}$O+$^{16}$O fusion cross sections and the
branching to the different channels. For the study of nucleosynthesis in massive stars, the AsFiN collaboration will continue the investigation of the reactions involved in the Ne-Na and Mg-Al cycles and,
with the forthcoming noble gases ion source, complete the investigation of the
$^{22}$Ne($\alpha$,n)$^{25}$Mg reaction.
\item Radioactive nuclei play a major role in astrophysics, especially in the case of
explosive scenarios.
These include long-lived species, having half-lives so large to allow for the production
of a target for a sputtering source, such as $^{10}$Be (T$_{1/2}=1.5\cdot10^6$~y),
$^{26}$Al$_{g.s.}$ (T$_{1/2}=7.2\cdot10^5$~y), $^{36}$Cl (T$_{1/2}=3.0\cdot10^5$~y), and short
lived species for which ISOL or in-flight RIB facilities are necessary for their use as
beam. AsFiN, in collaboration with the LNS staff and international laboratories
such as PSI (Switzerland) or ATOMKI (Hungary), has positively tested the possibility
to accelerate long-lived species and has proposed the indirect measurements of the
$^{10}$Be($\alpha$,n)$^{13}$C, of the $^{26}$Al(n,$\alpha$)$^{23}$Na and of the
$^{26}$Al(n,p)$^{26}$Mg reactions using the THM at LNS (in the case of $^{26}$Al
for the ground state only), of great importance in the understanding of the solar system
formation.
\item In the case of short-lived radioactive nuclei, AsFiN aims at investigating mainly
reactions influencing our understanding of the sources targeted by multimessenger
astronomy. For instance, the investigation of Type I X-ray bursts would help to
constrain neutron star models.
Among the reactions of interest, AsFiN considers the $^{14}$O($\alpha$,p)$^{17}$F
and the $^{18}$Ne($\alpha$,p)$^{21}$Na that could be the onset of a possible route that
breaks out from the HCNO cycles, and the $^{34}$Ar($\alpha$,p)$^{37}$K, since $^{34}$Ar
is a waiting point in X-ray bursts nucleosynthesis. As long as SPES will not be available, the AsFiN collaboration plans to perform these measurements at LNS using the in Flight Radioactive Ion Beams (FRIBs) facility at LNS \cite{RUS18} and at other RIBs facilities such as CRIB-RIKEN and MARS-TAMU; then the research
activity will be shifted on RIBs at SPES, owing to the foreseen intensities and beam quality. In particular, there is an accepted letter of intent about measurements involving $^{26}{\rm Al}^{m}$ (isomeric state
at 228 keV, T$_{1/2}=6.35$~s) beam, for studying the $^{26}$Al$^m$(n,$\alpha$)$^{23}$Na and
the $^{26}$Al$^m$(n,p)$^{26}$Mg reactions by means of the THM, using a deuteron to transfer
a neutron. $^{26}{\rm Al}^{m}$ should be the first beam
delivered by SPES using the SiC production target. At SPES, we also plan to measure
a number of (n,p), (n,$\alpha$) and (n,$\gamma$) reactions of interest for the
r-process using the THM and the ANC techniques. Finally, a developing activity making
use of radioactive nuclei is the investigation of $\beta$-delayed neutron emission
\cite{MUM16}. Indeed, during the r-process after freeze-out some nuclei may be populated
in neutron unbound states following $\beta$-decay.
\item Regarding fusion reactions in laser-induced hot plasmas, the physical cases investigated so far will be extended. Next
scheduled measurements are the $^{10}$B(p,$\alpha$)$^{7}$Be and the $^{11}$B(p,$\alpha$)$^{8}$Be
reactions, of interest for the electron screening problem and for aneutronic fusion in
terrestrial fusion reactors.
 \item The availability of high-resolution and high-intensity $\gamma$-beam has triggered
the interest for a number of direct and indirect measurements. Regarding direct measurements, they play a substantial role
in stellar advanced evolutionary stages, such as silicon burning before the onset of core collapse
supernova or the p-process.\\
Scheduled measurements are the $^{24}$Mg($\gamma,\alpha$)$^{20}$Ne governing the downward
flow from $^{24}$Mg to $\alpha$'s, the ($\gamma$,p) reactions on $^{74}$Se, $^{78}$Kr, $
^{84}$Sr, $^{92}$Mo, and $^{96}$Ru, as well as the the $^{96}$Ru($\gamma$,$\alpha$)$^{92}$Mo
reaction.
\end{itemize}

\mbox{} \newline
The driving idea of the AsFiN experimental program is to devise methods and adopt
procedures to approach as much as possible the same conditions as in
the astrophysical scenarios under investigation. To this purpose, AsFiN makes use of
either indirect techniques or resort to novel experimental apparatuses, such as
high-power lasers and $\gamma$-beams. This way, AsFiN could get a deeper look inside most of the astrophysical phenomena investigated so far. For instance, in \cite{TUM18}, the $^{12}$C+$^{12}$C could be explored for the first time in the energy range of astrophysical interest. This is a significant step forward in comparison with existing direct data \cite{BEC81,AGU06,MAR73,HIG77,KET80}. Moreover, at odds with direct measurements, indirect methods as the THM are the only ones which allow for the extraction of electron screening free astrophysical factors.

\mbox{} \newline
The activity performed by AsFiN is leading the field in the context of indirect
measurements for applications to nuclear astrophysics. THM and ANC give more
and more reliable data in the Gamow energy range to the astrophysics community
in many cosmic scenarios. Synergy is required with direct measurements
especially for the normalization procedure at higher energies. Moreover in
many other contexts, e.g. in explosive scenarios where the interaction
between neutrons and exotic beams has to be studied at astrophysical
energies, indirect methods give a unique possibility. Even large and
important collaborations and facilities like LUNA, JUNA, n$\_$TOF etc.
have limited possibilities for those scenarios.

\subsection{ERNA}

\begin{figure}[!htbp]
\begin{center}
\includegraphics[width=.95\textwidth]{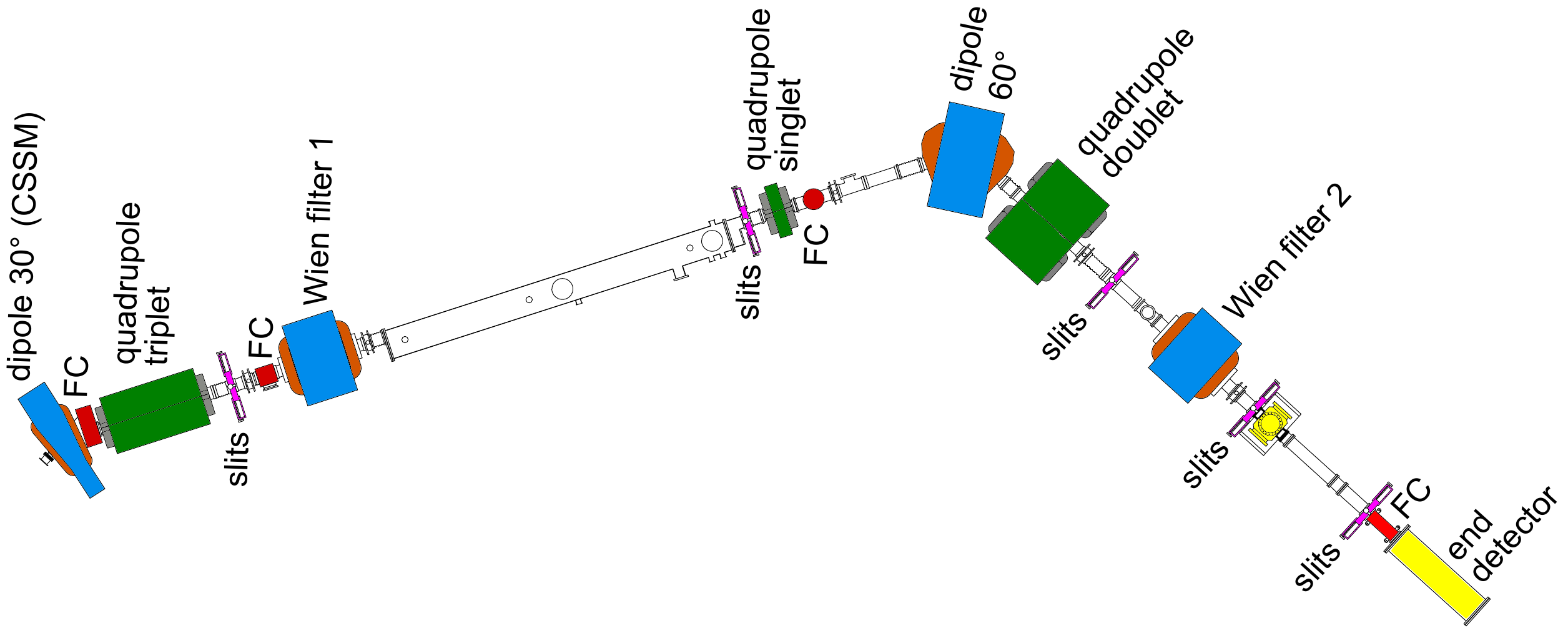}
\caption{Layout of the ERNA recoil separator at the CIRCE Laboratory.}
\label{fig:separator}
\end{center}
\end{figure}

The European Recoil Separator for Nuclear Astrophysics (ERNA) is a tool for the measurement of the total cross section of capture reactions of astrophysical interest. It was originally commissioned at the Dynamitron Tandem Laboratorium of the Ruhr-Universit\"at Bochum, Germany \cite{Rogalla1999a,Rogalla1999b,Rogalla2003,Gialanella2004,Schuermann2004}. Its design was focussed to the measurement of the $^{12}\mathrm{C}(\alpha,\gamma)^{16}\mathrm{O}$ in a wide energy range \cite{Rogalla1999a}. The excellent results on the total cross section of this reaction \cite{Schuermann2005,circe2013} were complemented with measurements of the cascade transitions \cite{Schuermann2011}. Significant advances were also brought to the understanding of the $^3\mathrm{He}(\alpha,\gamma)^7\mathrm{Be}$ process \cite{DiLeva2008,DiLeva2009,DiLeva2016}.\\
In 2009 ERNA was transferred to the Center for Isotopic Research on Cultural and Environmental heritage (CIRCE) of the Universit\`a della Campania/Innova. This gave the opportunity for a significant upgrade of the apparatus. The separator in its present layout is shown in Fig. \ref{fig:separator}.

In addition to the measurements with the recoil separator, the scientific activities of the ERNA Collaboration include also measurements with an array of double stage detectors (GASTLY), and the mass spectrometry of meteoritic samples, as well as other measurements that make use of the intense $^7\mathrm{Be}$ beam developed for the measurement of the $^{7}\mathrm{Be}$(p,$\gamma$)$^{8}\mathrm{B}$ reaction.

\mbox{} \newline
Three main experimental techniques are exploited at ERNA:

\begin{itemize}
\item On line recoil mass separation: The recoil separator technique allows the total cross section measurement of radiative capture reaction cross sections through the direct detection of the recoil nuclei produced in the fusion process. Although simple in principle, the technical challenges involved in its setup are enormous.\\*
The reaction is initiated in a windowless gas target differentially pumped on both sides. A number of gas targets were developed for the different reaction studied: extended $^4\mathrm{He}$ \cite{Gialanella2004,Schuermann2004,DiLeva2017}, extended $^3\mathrm{He}$ with gas recirculation \cite{DiLeva2008},  extended $^1\mathrm{H}_2$ \cite{Schuermann2011} and supersonic jet $^4\mathrm{He}$ \cite{Rapagnani2017}.\\*
The recoils emerging from the target are transported, for a selected charge state, to the end detector, while the primary beam is suppressed by many orders of magnitude ($10^{10}$-$10^{15}$) allowing for direct identification and counting of the reaction products with either a Ionization Chamber Telescope or Time of Flight-Energy detection system. For each reaction under study the transport of the recoils through the separator is ensued by transporting a beam with the same rigidity expected for the recoils, with maximum angle expected for them to emerge from the target, and experimentally determining the separator elements fields.

\item Charged particle spectroscopy: The study of the $^{12}\mathrm{C}$+$^{12}\mathrm{C}$ fusion process was initiated long ago with a relatively simple setup \cite{Zickefoose2018}. An high intensity beam, $\sim10\,\mathrm{p}\mu\mathrm{A}$, impinging on an thick graphite target, while charged particles are detected in detector telescopes at backward angles. That study approached the lowest measurement energies, however the limitations of the approach appeared clear, some were connected to the detection setup, the alpha particles could not be observed due to limitations of the $\Delta E$ detectors used, and some were connected to the background due to the hydrogen contained in traces in the graphites used as targets.\\*
To improve the situation, a thorough study of the behaviour of the target and the contaminant hydrogen under high current ion beam bombardment was conducted \cite{Morales-Gallegos2014,Morales-Gallegos2015,Morales-Gallegos2018}.\\*
To proceed with measurements towards the astrophysically relevant energies the ERNA Collaboration has developed a detection array GAs-Silicon Two-Layer sYstem (GASTLY) hosted in a large volume scattering chamber. It has been designed to detect and identify low-energy light particles emitted in nuclear reactions of astrophysical interest. Devoted to the measurement of nano-barn cross-sections, the system is optimised for large solid angle coverage and for low-energy detection thresholds. The array consists of eight modules, each comprising an ionisation chamber ($\Delta E$) and a large area silicon strip detector ($E_{\mathrm{rest}}$). Its modularity and versatility allow for use in a variety of experiments for the detection and identification of light charged particles.

\item Isotopic ratio mass spectrometry: The $^{60}\mathrm{Fe}$ and the $^{107}\mathrm{Pd}$ present in the early solar system, $\beta^-$ decayed to $^{60}\mathrm{Ni}$ and $^{107}\mathrm{Ag}$, with half lives of 2.62\,My and 6.5\,My, respectively. The measurement of the ratios of Ni and Ag isotopes abundance in meteorite samples provides information on the stellar event that generated the solar system. Such measurements can be performed to high accuracy with the Multicollector-Inductively Coupled Plasma Mass Spectrometer (MC-ICP-MS) present at the CIRCE Laboratory. In fact, the Inductively Coupled Plasma (ICP) source has a very high extraction efficiency ($>40\%$), that in conjunction with the multicollector, allows measurements in high resolution $M/\Delta M\simeq400$. In addition, this methodology can be also applied to metalloids and non-metals, greatly extending the number of isotopes that can be potentially measured.
\end{itemize}
\mbox{} \newline
Several nuclear astrophysics measurements are currently ongoing  at ERNA:
\begin{itemize}
\item At the recoil separator, the cross section determination of $^{7}\mathrm{Be}$(p,$\gamma$)$^{8}\mathrm{B}$ was recently completed. The procedures and results of a test measurement are presented in \cite{Buompane2018}.\\*
After this measurement the extended $^1\mathrm{H}_2$ target was replaced by the $^4\mathrm{He}$ supersonic jet gas target, actually starting the measurement campaign of the $^{12}\mathrm{C}(\alpha,\gamma)^{16}\mathrm{O}$, for which some of the ancillary measurements have already been performed (separator tuning, post-stripper characterization, charge state probability, etc.).\\
In parallel also the $^{14}\mathrm{N}(\alpha,\gamma)^{18}\mathrm{F}$ and the $^{15}\mathrm{N}(\alpha,\gamma)^{19}\mathrm{F}$ will be also investigated. The first one has already been studied with the extended $^4\mathrm{He}$ target \cite{DiLeva2017}. The narrow resonances below 1\,MeV will be measured including the most relevant at 364\,keV, known only through indirect methods, although this might be very challenging due to the extremely low expected counting rate ($\sim1\,$recoil/day).\\
Since at CIRCE is available a very intense beam of $^{7}\mathrm{Be}$ with tandem quality, to complement the $^{7}\mathrm{Be}$(p,$\gamma$)$^{8}\mathrm{B}$ measurement, the elastic scattering process $^{7}\mathrm{Be}$(p,p)$^{7}\mathrm{Be}$ is also being studied.

\item After a long phase of detector development and target characterization, a first measurement campaign of the $^{12}\mathrm{C}$+$^{12}\mathrm{C}$ has been completed in the energy range $E_{\mathrm{c.m.}}=2.3-4.5\,\mathrm{MeV}$. Measurements toward $E_{\mathrm{c.m.}}=2.0\,\mathrm{MeV}$ are ongoing. Concurrently, thanks to a further detector development that allows the use of silicon strip detectors as $E_{\mathrm{rest}}$ detectors, angular distributions of the emitted charged particles are being measured at  $E_{\mathrm{c.m.}}\sim4\,\mathrm{MeV}$.

\item For what concerns the analysis of meteorites, the extraction procedures for both Ni and Ag have been developed and characterized using standard certified materials.\\*
The first samples directly measured are two ferritic meteorites, the Brenham and the Mineo pallasites. The first one is a meteorite already studied and is therefore used as a benchmark for the developed methodology. The only available sample of the Mineo pallasite is present at the Dipartimento di Fisica e Geologia of the Universit\`a di Perugia. The morphological and chemical characterization of this sample has been the preliminary step toward the isotopic abundance ratios measurements \cite{Zucchini2018}.
\end{itemize}
\mbox{} \newline
The future scientific program of ERNA will be detailed next year. It will include the study of the $^{16}\mathrm{O}(\alpha,\gamma)^{20}\mathrm{Ne}$ and
$^{33}\mathrm{S}$(p,$\gamma$)$^{34}\mathrm{Cl}$. It might still include the $^{12}\mathrm{C}(\alpha,\gamma)^{16}\mathrm{O}$ reaction, depending on the results of the present campaign.\\*
As for charged particle spectroscopy, the study of $^{12}\mathrm{C}$+$^{12}\mathrm{C}$ might be continued also in view of the recent results obtained with the Trojan Horse Method. Also the study of the $^{12}\mathrm{C}$+$^{16}\mathrm{O}$ fusion processes is being considered.\\*
The present studies of meteorites were concentrated on bulk material, in the near future it will be extended to inclusions, mostly olivines as regards the pallasites, and Calcium-Aluminum inclusions (CAI) in chondrite samples. The study of CAIs will require the development of protocols for the preparation and measurement of $^{26}\mathrm{Mg}$ and $^{41}\mathrm{K}$, relics of $^{26}\mathrm{Al}$ and $^{41}\mathrm{Ca}$, respectively.\\
In addition, ERNA plans to start some measurements of the $^{7}\mathrm{Be}$ half life, neutral and ionized, is different environments. Also the measurement of $^7\mathrm{Be}(\alpha,\alpha)^7\mathrm{Be}$ elastic scattering has been proposed by the Notre Dame University group and a feasibility study is undergoing.

\mbox{} \newline
Concerning the measurements with the recoil separator, the measurement of the total cross sections allows to put in evidence components difficult to be identified through $\gamma$-ray spectroscopy, e.g. non radiative and/or weak cascade transitions, as in the case of $^{12}\mathrm{C}(\alpha,\gamma)^{16}\mathrm{O}$ \cite{Schuermann2004,Schuermann2011}. In addition, the use of a technique with systematics quite different from previous measurements, can in general improve the accuracy of global analysis cross section determinations.\\*
More in detail, for the $^{12}\mathrm{C}(\alpha,\gamma)^{16}\mathrm{O}$ the measurement of the cross section down to $E_{\mathrm{c.m.}}\sim1\,$MeV, with the concurrent measurement of $\gamma$-rays and $e^+e^-$ pairs at selected energies, will allow for more precise and accurate extrapolations to astrophysical energies.\\*
In the $^{12}\mathrm{C}$+$^{12}\mathrm{C}$ study ERNA aims to reach energies around 2\,MeV, improving the statistical precision of the data. Concurrently the measurement of angular distributions at higher energies might help in better identifying the parameters of resonances involved in the fusion process.\\
As concerns the meteorite studies, there are not many groups measuring isotopic ratios, especially for Ni. Therefore, ERNA expects to contribute to lower systematic uncertainties through the inter-comparison with similar measurements. Moreover, samples not studied before will be measured.

\mbox{} \newline
Several recoil separators devoted to the measurement of cross section of astrophysical interest are present around the world, see e.g. \cite{Ruiz2014} for a review.\\
The DRAGON recoil separator at TRIUMF Vancouver, Canada, is the most active. Originally designed for the measurement of reactions involving radioactive ion beams, has also measured over the years a number of reactions with stable nuclei. Given the limited acceptance of this separator only a moderate number of reaction can be meaningfully measured there.\\
The Daresbury Recoil Separator at the Holifield Radioactive Ion Beam Facility of the Oak Ridge National Laboratory, TN USA, is operational since 1997, however only a limited number of proton capture reactions have been studied. Its status at present is unclear.\\
The St. George recoil separator has been recently installed at the University of Notre Dame, IL USA. Equipped with a $^4\mathrm{He}$ jet gas target, it is devoted to measurements of reactions involving stable isotopes. At present is still in commissioning phase.\\
At the Kiushu University Tandem Laboratory, Kiushu Japan, was present a separator optimized for the study of the $^{12}\mathrm{C}(\alpha,\gamma)^{16}\mathrm{O}$ reaction. The separator has been recently disassembled for laboratory rebuilding works. It is unclear whether it will be reassembled and if other reactions are in the scientific program.\\
A few projects have been announced regarding recoil separators for nuclear astrophysics, mostly related to new RIB facilities, however they might also be used with stable beams:\\*
The Separator for Capture Reactions (SECAR) at ReA3 (low-energy reaccelerated beams) at the Facility for Rare Isotope Beams (FRIB) at Michigan State University, foreseen completion 2020-2022,\\*
The KOrea Broad acceptance Recoil spectrometer and Apparatus (KOBRA) at the RAON Heavy Ion Accelerator complex, Daejeon, South Korea. Completion foreseen in 2021.\\
As for $^{12}\mathrm{C}$+$^{12}\mathrm{C}$, although there is large interest on this reaction, the only ongoing measurement we are aware of started a few years ago at the University of Notre Dame. However, only a conference proceeding was published so far on this experiment \cite{Fang2013}.

\subsection{LUNA}
\begin{figure}[h]
\begin{center}
\includegraphics[angle=0,width=0.8\textwidth]{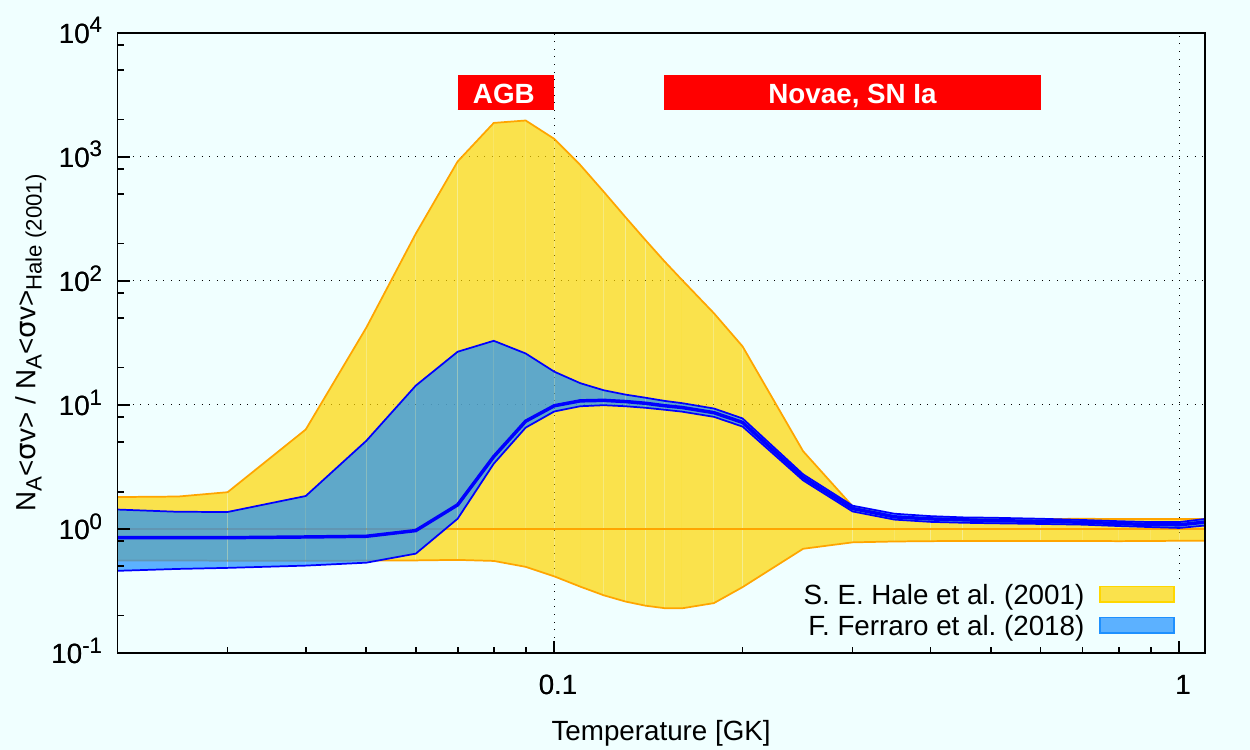}
\caption{Astrophysical reaction rate for the $^{22}$Ne(p,$\gamma$)$^{23}$Na reaction \cite{Ferraro-2018}, normalized to the rate from a previous indirect measurement \cite{Hale-2001}. The band widths show the uncertainties.}
\label{fig:luna-22Ne}
\end{center}
\end{figure}

Underground nuclear astrophysics was born twenty-seven years ago in the core of Gran Sasso, with the aim of measuring cross sections in the low energy range and of deriving reaction rates directly at stellar temperatures. LUNA (Laboratory for Underground Nuclear Astrophysics) started its activity as a pilot project with a 50 kV accelerator \cite{Greife94-NIMA} and it has been for about 25 years the only laboratory in the world running an accelerator deep underground, currently a 400 kV accelerator with hydrogen and helium beams \cite{Formicola03-NIMA}. The extremely low laboratory background has allowed for the first time nuclear physics experiments with very small count rates, down to a couple of events per month. Only in this way, the important reactions responsible for the hydrogen burning in the Sun could be studied down to the relevant stellar energies \cite{Costantini09-RPP, Broggini10-ARNPS}. In particular, the first direct determination of the cross section of $^{3}$He($^{3}$He,2p)$^{4}$He inside the solar Gamow peak definitively excluded a nuclear solution to the solar neutrino problem \cite{Bonetti99-PRL} while the LUNA experiment on the bottleneck reaction of the CNO cycle, $^{14}$N(p,$\gamma$)$^{15}$O, reduced by a factor of two the value of the CNO neutrino flux expected from the Sun \cite{Formicola04-PLB}.  More recently, LUNA shed light on the origin of one group of meteoritic stardust \cite{Lugaro17-NA} through a low-energy study of the $^{17}$O(p,$\alpha$)$^{14}$N \cite{Bruno16-PRL}. Moreover a direct measurement of the $^{22}$Ne(p,$\gamma$)$^{23}$Na reaction, taking part in the Ne-Na cycle, lead to the first direct observation of three low-energy resonances, with major impact on the astrophysical reaction rate \cite{Depalo-2016, Ferraro-2018} (see Fig. \ref{fig:luna-22Ne}). Full descriptions of LUNA and several results are given in two recent review papers \cite{Broggini18-PPNP, Cavanna18-IJMPA}.

\mbox{} \newline
The 400 kV current LUNA accelerator and the unique low-background conditions of the underground LNGS laboratory have been and still are the perfect blend for the study of most of the proton-capture reactions involved in the stellar H burning and in the Big Bang nucleosynthesis.\\
Among the reactions so far studied by the LUNA-400kV collaboration, there are various processes of the CNO cycle involving nitrogen and oxygen isotopes, i.e., $^{14}$N(p,$\gamma$), $^{15}$N(p,$\gamma$), $^{17}$O(p,$\alpha$), $^{17}$O(p,$\gamma$) and $^{18}$O(p,$\gamma$), and some of the Ne-Na and  Mg-Al cycle, i.e. $^{22}$Ne(p,$\gamma$) and $^{25}$Mg(p,$\gamma$). A scientific program based on the study of D(p,$\gamma$), $^{6}$Li(p,$\gamma$), $^{13}$C($\alpha$,n), $^{22}$Ne($\alpha$,$\gamma$) and $^{12,13}$C(p,$\gamma$) has been approved both by INFN-CSN3 and LNGS-Scientific Committee and it will extend to the whole 2019.\\
Theoretical calculations of the products of the primordial nucleosynthesis (also Big Bang nucleosynthesis, hereinafter BBN) provide important hints for cosmology and particle physics. The recent developments of experimental cosmology, such as, in particular, the precise determination of the CMB temperature fluctuations obtained by WMAP and Plank as well as the observation of high-redshift supernovae, revived the interest for BBN studies. In practice, different observables provide spots of the Universe at different epochs. In particular, LUNA-400kV has contributed to improve the predictions of the BBN by significantly reducing the uncertainties of the rates of the $^{3}$He($\alpha$,$\gamma$)$^{7}$Be and the D($\alpha$,$\gamma$)$^{6}$Li reactions. To complete this investigation, an accurate determination of the D(p,$\gamma$)$^{3}$He reaction rate is now missing and is the goal of the ongoing experiment at LUNA-400kV. This reaction mainly affects the primordial deuterium abundance. A precise determination of its rate at BBN energies is a necessary input to constrain baryon density and effective neutrino family number.\\
In view of the foreseen studies at the new LUNA MV facility, experiments on two processes that may be active in the stellar He-burning zones, when the slow neutron capture nucleosynthesis (s-process) takes place are also in progress. The first, the $^{13}$C($\alpha$,n)$^{16}$O reaction, is the most important neutron source in low-mass Asymptotic Giant Branch (AGB) stars and is responsible for the production of about half of the heavy isotopes (beyond iron) in nature. The second is the $^{22}$Ne($\alpha$,$\gamma$)$^{26}$Mg, which competes with the $^{22}$Ne($\alpha$,n)$^{25}$Mg reaction, another important neutron source for massive stars, and contributes to the synthesis of Mg isotopes.
The on-going program is completed by the study of the $^{6}$Li(p,$\gamma$)$^{7}$Be reaction to clarify some important nuclear physics issues and to improve the knowledge of the $^{3}$He($\alpha$,$\gamma$)$^{7}$Be process, one of the key branches of the pp chain.
Projects on other scientific cases which could be perfectly suited for cutting-edge experiments at LUNA-400kV are under discussion and will be collected in an organic proposal within the end of 2019.

\begin{figure}
\begin{center}
\includegraphics[angle=0,width=1.0\textwidth]{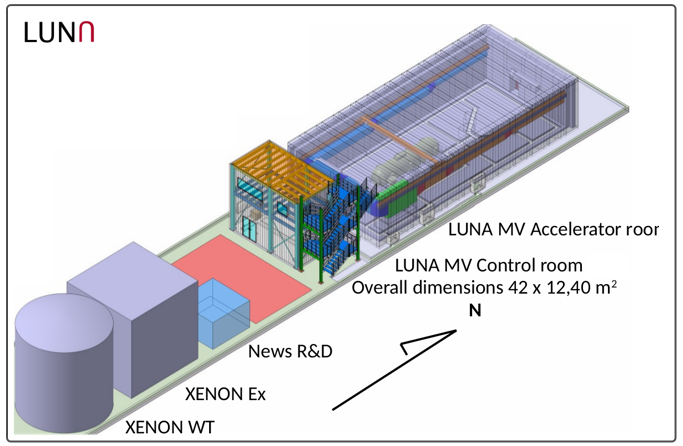}
\caption{Layout of the LUNA MV installation with the 3.5 MV accelerator in Hall B at LNGS (courtesy of Eng. L. Di Paolo).}
\label{fig:luna-mv}
\end{center}
\end{figure}

\mbox{} \newline
A beam of higher energy is required to extend the LUNA approach to study He and C burnings. The LUNA MV project has been developed to overcome such a limit with the new 3.5 MV single-ended accelerator to be installed in Gran Sasso at the beginning of 2019. The accelerator will provide hydrogen, helium and carbon (also doubly ionized) high current beams and it will be devoted to the study of those key reactions of helium and carbon burning that determine and shape both the evolution of massive stars towards their final fate and the nucleosynthesis of most of the elements in the Universe.\\
$^{12}$C+$^{12}$C and $^{12}$C($\alpha$,$\gamma$)$^{16}$O represent the ``Holy Grail'' of nuclear astrophysics and they are the most ambitious goals of this project. In particular, $^{12}$C+$^{12}$C will be the flagship reaction of the first 5 years proposal of LUNA MV. This reaction is the trigger of C burning. The temperature at which C burning takes place depends on its rate: the larger the rate, the lower the C-burning temperature. Since the temperature controls the nucleosynthesis processes, reliable estimations of all the yields produced by C burning, for example the weak component of the s-process which produces the elements between Fe and Sr, require the precise knowledge of the $^{12}$C+$^{12}$C rate. The $^{12}$C+$^{12}$C rate also determines the lower stellar mass limit for C ignition. This limit separates the progenitors of white dwarfs, novae and type Ia supernovae, from those of core-collapse supernovae, neutron stars, and stellar mass black holes. This mass limit also controls the estimations of the expected numbers of these objects in a given stellar population, which are required to answer crucial questions such as: how many neutrons stars are there in the Milky Way? How many double neutron stars are there in close binaries? And what is the expected merging rate?\\
Among the key processes for stellar nucleosynthesis, the sources of neutrons represent a longstanding and debated open problem \cite{Burbidge57-RMP, Cameron57-PASP}. Neutron-captures (slow or rapid, i.e., the s- or r- process, respectively) were early recognized as the most important mechanism to produce the elements heavier than iron. The identification of the astrophysical sites where these processes may operate requires the accurate knowledge of the efficiency of the possible neutron sources. Various reactions have been identified as promising neutron sources. Among them $^{13}$C($\alpha$,n)$^{16}$O and $^{22}$Ne($\alpha$,n)$^{25}$Mg represent the most favored candidates. This is because they operate from relatively low temperatures typical of He burning (100 - 300MK) and because $^{13}$C and $^{22}$Ne are relatively abundant nuclei in stellar interiors.\\
The $^{13}$C($\alpha$,n)$^{16}$O reaction operates in the He-burning shell of low-mass (less than 4 solar masses) AGB stars and it is the neutron source reaction that allows the creation of the bulk of the s-process elements such as Sr, Zr and the light rare earth elements in the Universe. The $^{22}$Ne($\alpha$,n)$^{25}$Mg reaction operates in the He-burning shell of high-mass (more than 4 solar masses) AGB stars and during the core-He burning and the shell-C burning of massive stars (more than 10 solar masses). Underground experiments with LUNA MV will allow us to gain a full understanding of these two reactions through the direct measurement of their cross sections in the energy range of astrophysical interest.\\
The LUNA MV facility will be installed at the north side of Hall B and will consist of an accelerator room with concrete walls and a further building hosting the control room and technical facilities including the cooling system, the electric power center, etc. (Fig. \ref{fig:luna-mv}). The concrete walls and ceiling (thickness of 80 cm) of the accelerator room serve as neutron shielding. Considering the worst case scenario for the operation of the LUNA MV facility of maximum neutron production rate of R$_{n}$ = 2 $\cdot$ 10$^{3}$ s$^{-1}$ with an energy E$_{n}$ = 5.6 MeV, validated Monte Carlo simulations determined a maximum neutron flux outside the shielding of $\phi_{max}$ = 5.70 $\cdot$ 10$^{-7}$ cm$^{-2}$ s$^{-1}$. This value is a factor of 5 lower than $\phi_{LNGS}$ = 3 $\cdot$ 10$^{-6}$ cm$^{-2}$ s$^{-1}$, the reference neutron background at LNGS. In addition, the energy distribution of the neutrons produced by LUNA MV just outside the shielding is very similar to that of the natural background at LNGS.\\
The LUNA MV accelerator is an inline Cockcroft Walton accelerator designed and constructed by  High Voltage Engineering Europe (HVEE). The machine will cover a Terminal Voltage (TV) range from 0.2 to 3.5 MV and will deliver ion beams of H$^{+}$, $^{4}$He$^{+}$, $^{12}$C$^{+}$ and $^{12}$C$^{++}$ in the energy range from 0.350 to 7 MeV into two different beam lines via a 35$^{\circ}$ switching analyzing magnet. The two independent targets will be located at 2 m distance from the analyzing magnet. Details of the characteristics of the machine can be found in Table \ref{tab:luna-mv}.

\begin{table}[h]
\begin{center}
\begin{tabular}{ l l r }
\hline
\multicolumn{3}{p{13cm}}{\footnotesize{Expected ion beam currents transported through an aperture (length 40 mm, diameter 5 mm) located at the target position}} \\
\hline
$^{1}$H$^{+}$ & TV: 0.3 - 0.5 MV & 500 e$\mu$A \\
 & TV: 0.5 - 3.5 MV & 1000 e$\mu$A \\
$^{4}$He$^{+}$ & TV: 0.3 - 0.5 MV & 300 e$\mu$A \\
 & TV: 0.5 - 3.5 MV & 500 e$\mu$A \\
$^{12}$C$^{+}$ & TV: 0.3 - 0.5 MV & 100 e$\mu$A \\
 & TV: 0.5 - 3.5 MV & 150 e$\mu$A \\
$^{12}$C$^{2+}$ & TV: 0.5 - 3.5 MV & 100 e$\mu$A \\
\hline
\multicolumn{3}{c}{Other ion beam parameters} \\
\hline
{Beam current stability} & over 1 h & 5$\%$\\
 & over 1 min & 2$\%$\\
Beam energy stability over 1 h & & \\
\hspace{2pt} whichever higher & & $10^{-5} \times$ TV or 20 V\\
Beam energy reproducibility & & \\
\hspace{2pt} whichever higher  & & $10^{-4} \times$ TV or 50 V\\
\hline
\multicolumn{3}{c}{Operational details} \\
\hline
Ion species change-over duration & & $\leq$ 30 min\\
Intervention-free operation & & $>$ 24 h\\
Interruption time after maximum & & \\
\hspace{2pt} intervention-free operation& & $<$ 45 min\\
Servicing interval & & 700 h\\
Annual operation capabilities & & 7400 h \\
\hline
\end{tabular}
\caption{Main features of the LUNA MV accelerator \cite{Sen-2018}}
\label{tab:luna-mv}
\end{center}
\end{table}%

The delivery of the accelerator to LNGS is scheduled by the end of this year. The six months installation and commissioning phase at LNGS will start directly after installation and is under the responsibility of HVEE. The scientific life of the new facility will be of 25-30 years just considering the possible experiments in nuclear astrophysics and can go well beyond if including applications to other fields even outside the fundamental research frame. The nuclear astrophysics program includes the direct measurement of the cross section of the $^{12}$C+$^{12}$C, $^{13}$C($\alpha$,n) and $^{22}$Ne($\alpha$,n) reactions and a renewed study of the $^{14}$N(p,$\gamma$) at energies higher than those previously explored at LUNA-400kV (this experiment will also be partially used in the LUNA MV commissioning phase). In a successive phase,  $^{12}$C($\alpha$,$\gamma$) should also be studied with a dedicated set-up. The LNGS-Scientific Committee has thoroughly evaluated the proposal submitted by the LUNA-MV collaboration. The LNGS-SC recommended
to grant a 5-year priority to the LUNA-MV Collaboration in a scheme in which the LUNA-MV machine
is operated by LNGS as an open facility.

\mbox{} \newline
The core of the LUNA mission is to provide solutions or clear steps forward in the knowledge of nuclear processes relevant in astrophysical scenarios. The scientific cases selected for experiments at LUNA-400kV and LUNA MV will follow this basic criterion. For instance, a very recent experiment carried out with the Trojan Horse Method, claimed the existence of several resonances in $^{12}$C+$^{12}$C  in the so far unexplored region at E$_{cm}$ 1- 2.1 MeV \cite{TUM18}. This observation implies a significant increase in the reaction rate and a corresponding lowering of the mass limit of stars for which carbon-burning can ignite. An independent verification of such important result is definitively needed and LUNA MV is the sole laboratory in the world where a low-energy direct experiment can be performed. Actually, the $^{12}$C+$^{12}$C experiment will be the main goal/task in the first years of life of the new underground facility.

\mbox{} \newline
The LUNA achievements have motivated the proposals for two similar facilities currently under construction in the Republic of China (JUNA project) and in the United States (CASPAR project). In the first case, intense p and He beams (project target: 10 mA) will be accelerated up to 400 kV (deployment within 2019) while the American project is based on a refurbished 1 MV accelerator which has been put in operation in 2018 and that can deliver p and He beams with intensities in the range of 0.1 mA. The scientific cases of both facilities are partially overlapping with the LUNA MV program in particular for the processes which can be studied without a carbon beam (available at LUNA MV only). A shallow underground accelerator has been also installed in Dresden (Felsenkeller project): in this case, the background level is much less favorable, however, the availability of a carbon beam must be underlined. In short: LUNA is no more the sole underground nuclear astrophysics facility and the next decades will develop in a framework of strong international competition.

\subsection{n$\_$TOF}

\begin{figure}[!h]
\centering
\includegraphics[width=12cm,clip]{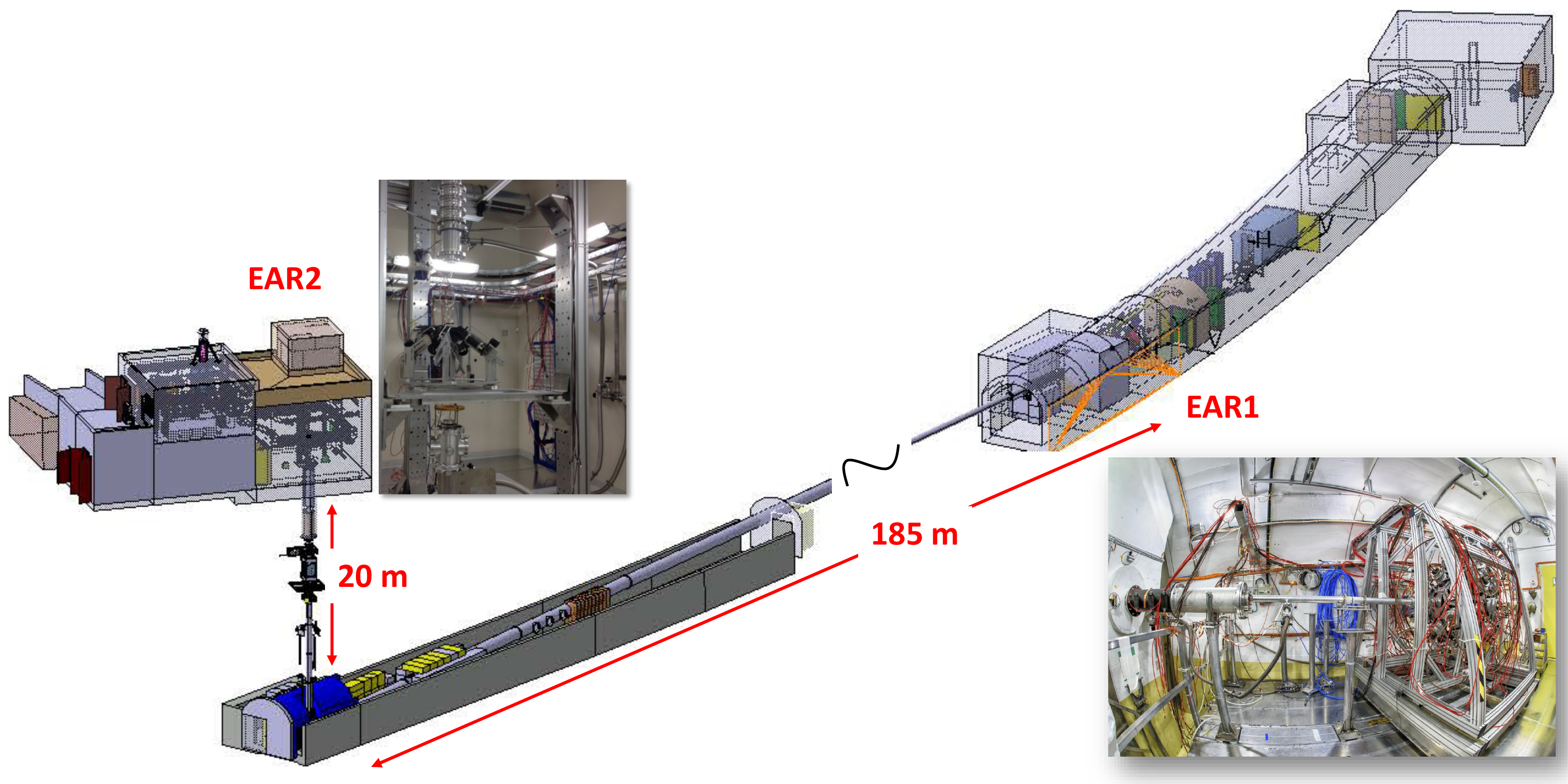}
\caption{Schematic view of the n$\_$TOF facility beam lines and experimental areas. The Experimental Area 1 (EAR1) is horizontal to the beam line with a flight path of 185 m, while the Experimental Area 2 (EAR2) is vertical to the beam line and has a flight path of 20 m.}
\label{fig1}     
\end{figure}

The n$\_$TOF facility was proposed and built at CERN with the aim of addressing the needs of new data of interest for nuclear astrophysics, as well as for basic nuclear physics and for technological applications \cite{Abbondanno-2005}. Neutrons are produced by spallation of a pulsed proton beam of 20 GeV/c momentum from the Proton Synchrotron (PS) impinging on a water-cooled Pb target. The high peak current of the PS proton beam, combined with the high energy, results in a very intense neutron source, and in a very wide energy spectrum. The use of demineralized or borated water ensures additional moderation of the neutron energy, leading to an almost isolethargic spectrum. For the first thirteen years of operation, only one neutron beam line, in the horizontal direction, was available, with the experimental area (EAR1) located at 185 m distance from the spallation target (see Fig. \ref{fig1}). In this area, the neutron beam is characterized by an instantaneous intensity of 10$^{6}$ neutrons/bunch, an energy spectrum extending over almost eleven orders of magnitude, from 25 meV to 1 GeV, and a high energy resolution ($\Delta$E/E $<10^{-3}$ in most of the energy range). The combination of the high luminosity and resolution makes the neutron beam in EAR1 very convenient for measurements of capture cross sections of interest for nuclear astrophysics, involving in particular unstable s-process branching point isotopes.

In 2014, a second beam line in the vertical direction was completed at n$\_$TOF \cite{Weiss-2015}. The corresponding experimental area (EAR2) is located at 20 m distance from the spallation target. Compared with EAR1, the neutron beam in the new area has a much higher intensity ($>10^{7}$ neutrons/bunch), at the expenses of a worse energy resolution and slightly smaller energy range. The two experimental areas are somewhat complementary: while the characteristics of the neutron beam in EAR1 are ideal for high resolution measurements of neutron capture reactions on stable or long-lived radioactive isotopes (t$_{1/2}>$100 y), the much higher flux in EAR2 allows one to perform measurements on radioisotopes with short half-life, on samples available in small (sub-mg) mass, on reactions with low cross section, or on all of these at the same time. In fact, for radioactive isotopes the combination of the higher flux and shorter time-of-flight results in an increase of the signal-to-background ratio of more than two orders of magnitude, relative to EAR1, when considering the background related to the radioactivity of the sample. As a consequence, the neutron beam in EAR2 allows challenging measurements on isotopes with half-life as short as a few months. This was indeed the case for the measurement of neutron-induced reactions on $^{7}$Be of interest for BBN \cite{Barbagallo-2016,Damone-2016}.\\
The extremely high luminosity of the neutron beam in EAR2 is particularly convenient for measurements of neutron-induced fission cross sections on short-lived actinides. Such measurements are of key importance for the correct modelling of the fission recycling in r-process nucleosynthesis.\\
Apart from the unique features of the neutron beams in the two experimental areas, research in nuclear astrophysics at n$\_$TOF can benefit from state of the art detection and acquisition systems \cite{Abbondanno-2008}. In measurements of neutron capture reactions, a large step forward in the accuracy of the results has resulted from the development of new $\gamma$-ray detectors characterized by a very low neutron sensitivity and from a specifically designed data acquisition system \cite{Abbondanno-2008}.
For fission measurements, a big improvement in the quality of the data has resulted from the use of Micromegas detectors and of ultra-fast Parallel Plate Avalanche Counters specifically designed for n\_TOF.\\

\mbox{} \newline
All the elements heavier than iron are synthesized by neutron induced reactions in the s-process or in the r-process.
Both, the s- and r-process, are based on neutron capture reactions. In particular, for the s-process there is a direct correlation between the neutron capture cross section, $\sigma(n,\gamma$) and the observed abundance of a given isotope. This correlation makes it necessary to obtain neutron capture cross section data of all isotopes along the valley of $\beta$-stability, a program which has been going on since the early days of nuclear astrophysics.
\\Recently, this classic picture just described has been enriched by the enormous progress made by astronomical observations as well as by important developments in understanding and modeling stellar evolution. In particular, not only the need for neutron capture cross sections is enhanced by these recent developments, but the necessity of high accuracy is evident. The direct impact of accurate neutron capture cross sections on the nucleosynthesis modeling can be tested with measurements of elemental abundances in low-metallicity stars as well as in isotopic abundances measured for SiC grains embedded in meteoritic material.
\\Particularly unique features of s-process nucleosynthesis are branchings in the reaction path. Since typical neutron capture times are of the order of one year, the majority of unstable isotopes encountered by the neutron capture chain decays so fast that neutron capture becomes negligible. A number of isotopes, however, exhibit half lives comparable with the neutron capture time. The resulting competition gives rise to a branching of the reaction path, a local phenomenon that involves usually no more than eight isotopes before the branching is closed and the capture path continues as a single reaction chain.
\\Detailed analyses of such branchings are so fascinating because the evolving abundance patterns reflect the physical conditions at the stellar site of the s-process. In the simplest case, this is the stellar neutron flux but there are branchings that are strongly determined by temperature, or even by the convective motions in the deep stellar interior. Obviously, this type of information represents stringent tests for stellar models of the AGB phase, which is known to be the stage when the s-process operates. A prominent example of a branching point is $^{151}$Sm, which determines the reaction flow towards the s-only isotope $^{152}$Gd. The neutron capture cross section of $^{151}$Sm (t$_{1/2}$ = 93 years) has been successfully measured at n$\_$TOF with an accuracy of better than 5$\%$ \cite{Abbondanno-2004}\cite{Marrone-2005}. Such measurements are difficult because the radioactivity of the sample may cause excessive backgrounds in the detectors or may be unacceptable because of standard safety limits. Both problems suggest that sufficiently sensitive techniques are required, which can tolerate the use of extremely small samples.
\\Since the s-process component can be reliably determined on the basis of accurate (n,$\gamma$) cross sections, the r-process counterpart is obtained by the difference with the observed abundances
\begin{equation}
N_{r-process}=N_{star}-N_{s-process}
\label{eq:uno}
\end{equation}
Recent spectral analyses of very old stars have shown that they contain essentially only r-process material and that the abundances of all elements heavier than Ba scale precisely as in the solar r-process pattern. Below Ba this perfect agreement appears strongly disturbed. This could either indicate the operation of two independent r-processes or could simply be due to uncertain s-process abundances, since the cross sections in this mass region exhibit large uncertainties.
\\A huge step forward in this matter has resulted from the recent multi-messenger observation of a NS-NS merger in August 2017. The simultaneous detection of gravitational waves, gamma-ray burst and electromagnetic radiation in the optical and near-optical range has unequivocally demonstrated that NS-NS merger are an important, if not the most important, site for r-process nucleosynthesis.

\mbox{} \newline
Based on these considerations, an experimental program to be performed at n$\_$TOF has been defined. This program includes capture cross section measurements on the stable isotopes of Fe, Ni, Zn and Se \cite{Lederer-2014,Giubrone-2014,Massimi-2017}. These elements are produced by s-process nucleosynthesis in massive stars. In a number of cases, the respective neutron capture cross sections act as bottle-necks in the s-process reaction flow, thus strongly affecting the production of the elements with higher masses. In addition to the stable isotopes, measurements on some radioactive targets in the same mass region can be performed at n$\_$TOF. Examples of branchings related to the weak s-process are $^{63}$Ni (t$_{1/2}$ = 100 years) \cite{Lederer-2013} and $^{79}$Se (t$_{1/2}$ = 6.3 $\cdot$ 10$^{4}$ years).
\\The second group of isotopes of interest consists of the stable isotopes of Mo, Ru and Pd. Their neutron capture cross sections are required for two main objectives: for the interpretation of the isotopic patterns in pre-solar SiC grains and for the reliable determination of the r-process abundances in the critical mass region between Zr and Ba. SiC grains are found in meteorites and originate from circumstellar clouds of AGB stars, which are highly enriched in fresh s-process material produced by these stars. These grains are so refractory that they survived the high temperatures during the collapse of the proto-solar cloud. Hence, their composition represents the most direct information on the s-process efficiency of individual stars, an important constraint for current s-process models.
\\The third group of measurements concerns unstable branch point isotopes related to the main s-process component. So far, the feasibility of such measurements has been demonstrated at n$\_$TOF at the examples of $^{151}$Sm (t$_{1/2}$ = 93 yr) and $^{93}$Zr (t$_{1/2}$ = 1.5 $\cdot$ 10$^{6}$ yr) \cite{Tagliente-2013} and should be extended to cases like $^{134}$Cs (t$_{1/2}$ = 2.06 yr).
Other unstable isotopes which could be considered for new measurements are: $^{85}$Kr, $^{107}$Pd, $^{135}$Cs and $^{185}$W.\\
A fourth group of neutron capture data fundamental for a detailed knowledge of the s-process are those of s-only isotopes and of magic nuclei. For instance, at n$\_$TOF data on $^{154}$Gd, $^{88}$Sr, $^{89}$Y and $^{140}$Ce are currently under analysis. In the future, other challenging measurements are planned at n$\_$TOF, as that of the neutron capture on $^{138}$Ba, which has one of the smallest neutron cross sections in the isotopic table.
\\The $^{13}$C($\alpha$,n)$^{16}$O reaction is the main neutron source in low mass AGB stars. When the direct approach to the study of ($\alpha$,n) reactions is particularly difficult, indirect or inverse reactions are considered as a valid method for constraining the reaction cross sections of astrophysical interest. Therefore, the challenging $^{13}$C($\alpha$,n)$^{16}$O  measurement can benefit from experimental information from the time-reversed reaction $^{16}$O(n,$\alpha$)$^{13}$C, which can provide information on the orbital angular momenta of the excited states.
\\Neutron induced reactions on light nuclei may be of key importance for heavy element abundances produced in the s-process, as they may act as neutron poison. The most important poisons for the s-process are the $^{14}$N(n,p)$^{14}$C and the $^{26}$Al(n,p)$^{26}$Mg reactions. The first works in radiative conditions, when neutrons are released by the $^{13}$C($\alpha$,n)$^{16}$O reactions, and it is able to shape the whole s-process distributions. The latter operates in a convective environment during the neutron burst provided by the $^{22}$Ne($\alpha$,n)$^{25}$Mg reaction and modifies isotopic compositions close to branchings along the s-process path.

\mbox{} \newline
Nuclear reactions responsible for the $^{7}$Be creation and destruction during Big Bang nucleosynthesis (BBN) play the key role in the determination of the resulting primordial abundance of $^{7}$Li. Current standard BBN models predict a $^{7}$Li abundance which is a factor of 2-3 larger than what can be determined by astronomical observations. A neutron channel which could enhance the destruction rate of $^{7}$Be during BBN has been the subject of recent research activities at n$\_$TOF.
\\The $^{7}$Be(n,$\alpha$)$^{4}$He reaction has been measured for the first time in a wide neutron energy range \cite{Barbagallo-2016}. A second reaction channel, the $^{7}$Be(n,p)$^{7}$Li has been explored \cite{Damone-2016}, again extending the reaction cross section data to a wider range. The estimate of the $^{7}$Be destruction rate based on the new results yields a decrease of the predicted cosmological Lithium abundance of 10$\%$, insufficient to provide a viable solution to the cosmological lithium problem. The two n$\_$TOF measurements of (n,$\alpha$) and (n,p)  cross sections on $^{7}$Be finally ruled out neutron-induced reactions as a potential explanation of the cosmological lithium problem, leaving all alternative physics and astronomical scenarios still open.

\mbox{} \newline
In the scenario of NSM, the large number of free neutrons per seed nuclei (of the order of a few hundred) leads to the production of heavy fissioning nuclei. As a consequence, neutron-induced fission reactions and, to a lesser extent, spontaneous fission play a fundamental role in the process, by recycling matter during neutron irradiation. Such process, referred to as fission recycling has an important effect on shaping the abundance distribution of r-nuclei in the mass region between A=110 and 170, as well as in determining the residual production of heavy nuclei, in particular Pb, Bi, Th and U. Finally, fission processes during r-process nucleosynthesis are expected to contribute to the heating of the material, due to the large energy released (the so-called kilonova event).
\\The effect of fission recycling depends on the fission probabilities of a large number of, mostly neutron-rich, nuclei. For the vast majority of these nuclei one can only rely on model prediction, which at present are highly uncertain. Various models have been developed and are currently available, but at present the prediction power of these models is still highly unsatisfactory. As an example, a comparison of the fission cross section estimated on the basis of these nuclear models shows discrepancies of up to a factor of ten relative to experimental cross sections, and it is reasonable to expect that the predictive power gets much worse for exotic nuclei. More work is therefore necessary to improve the situation towards more coherent and reliable fission models.
To this end, a great help could come from new experimental data on actinides that have not been measured so far, or whose experimental cross section is affected by a large uncertainty. In this respect, high-luminosity neutron beams that have just become available or will come on line in the next few years will certainly offer the unique opportunity to collect precious data on short-lived actinides of interest for nuclear astrophysics. This is the case, in particular, of the high-flux experimental area EAR2 of the n$\_$TOF facility at CERN.

\mbox{} \newline
In all cases discussed above, the main problem is related to the large uncertainty in the cross section data, that often represent the main limitation towards the refinement of theoretical models of nucleosynthesis and stellar evolution. The uncertainty in the nuclear input in turn reflects the lack of reliable experimental data.  In some cases, measurements have been performed long time ago, with inadequate detector systems. As an example, neutron capture reactions have been measured in the past with detectors affected by a non-negligible neutron sensitivity, resulting in a large background and, consequently, in unrecognized systematic errors. For radioactive isotopes, the natural radioactivity of the sample represents another important source of background in capture measurements, leading to poor quality results. In fission measurements, the strong background related to the $\alpha$-activity of the sample has often hindered the possibility to obtain accurate data, or even made the measurement impossible. Finally, the small amount of sample material available for some isotopes has made neutron measurements difficult, or has led to highly uncertain results.
All these limitations have been mostly overcome at n\_TOF thanks on the one hand to the high luminosity of the neutron beam, in particular in the second experimental area, and on the other hand to the optimization of the detection systems. Another improvement is related to the high resolution of the neutron beam in the first experimental area, which allows one to better identify and characterize resonances in the cross section. Finally, the wide energy range of the n\_TOF neutron beam, extending over 10 orders of magnitude, allows to cover in a single measurement the entire energy region of interest for nuclear astrophysics, for a wide range of stellar temperatures, and results in a considerable reduction of normalization problems.
All these features are expected to lead to sizable improvements over previous results, in terms of accuracy, resolution and energy range covered. Most importantly, several measurements that could not be performed in the past have now become feasible, so that data can be collected for the first time ever.

\mbox{} \newline
While a few high-flux activation facilities are now becoming available (such as SARAF in Israel, of the future FRANZ facility in Germany), n$\_$TOF is predicted to continue playing a fundamental role for the determination of energy-differential cross sections in the mid-term future (the next ten years), as even new facilities currently under construction (such as NFS in GANIL, France) will not be able to meet the unique combination of high luminosity, wide energy range, high resolution and low background that characterizes the n$\_$TOF neutron beams. These features are fundamental to drastically improve the accuracy on the cross section for a variety of neutron-induced reactions of interest for nuclear astrophysics. Most importantly, they will allow to perform challenging new measurements, not possible up to now, for example on short-lived radioactive isotopes. To this end, the possibility to irradiate samples produced by implantation of radioactive beams could result in a number of unique measurements. The viability of this technique has just been successfully demonstrated at CERN in the measurement of the $^{7}$Be(n,p)$^{7}$Li reaction of interest for BBN, with a $^{7}$Be sample produced at the Isotope Separator On Line DEvice (ISOLDE). In the landscape of the present and future neutron facilities worldwide, the ISOLDE/n\_TOF combination represents a unicum that will hardly be met in other laboratories.

\subsection{ALICE}

The measurements of astrophysical interest that can be carried out by the ALICE experiment~\cite{Aamodt:2008zz} using proton--proton and heavy-ion collisions are mainly those related
to the production of light nuclei and hypernuclei on the one hand, and to the interactions between pairs of baryons (two protons, a proton and a hyperon, and two hyperons) on the other hand.
These measurements can provide insight on the production of light nuclei and hypernuclei
in the early Universe and in cosmic-ray interactions, also in connection with dark matter searches.
The possible impact of these studies for the equation of state of compact stellar objects
with high baryon density cores, either non-strange or strange, is presented in section 7.

\mbox{} \newline
The production of protons and stable light nuclei is measured in ALICE in the central rapidity region using track reconstruction and heavy-particle identification
through ionization energy loss in the Time Projection Chamber gas and time-of-flight from the interaction point to the Time Of Flight detector.
The production of hyperons and hypernuclei is measured by reconstructing their weak-decay secondary vertices in the Time Projection Chamber and in the Inner Tracking System.

\mbox{} \newline
The measurement of light nuclei production in pp collisions at ultra-relativistic energies has an impact on cosmology.
Big Bang nucleosynthesis is the dominant natural source of deuterons and, in the absence of baryogenesis, one could assume that the same holds for anti-deuterons. These anti-nuclei and even heavier anti-nuclei can also be produced in pp and pA collisions in interstellar space, representing a background source in the searches for segregated primordial antimatter and dark matter~\cite{Donato:1999gy}.
However, experimental information on anti-d and anti-$^3$He production is scarce and, for the most part, limited to AA or pA collisions.
ALICE has recently measured the production of nuclei and anti-nuclei in pp collisions at different energies, including for the first time anti-nuclei heavier than anti-deuteron~\cite{Acharya:2017fvb}.
The ALICE results show an increasing trend in the (anti-)d/(anti-)p ratio with charged particle multiplicity. Furthermore the values reported in central AA collisions are in agreement with a thermal model description while the highest d/p ratio measured in pp collisions by ALICE is about half the thermal model value; therefore, a thermal-statistical description is disfavored in pp collisions at these low average charged particle multiplicities.
In addition the ALICE data are well described by QCD-inspired event generators when a coalescence-based afterburner that also takes into account the momentum correlations between nucleons is included.

\mbox{} \newline
The data samples of pp collisions that will be collected during LHC Run-2 and Run-3 will be larger by factor 10 and 100, respectively, with respect to that used in~\cite{Acharya:2018gyz}. These
will enable to constraint the phase space spanned by the effective range and scattering
length of the $\Lambda$-p and $\Lambda$-$\Lambda$ interaction and to
extend the measurement also to $\Sigma$, $\Xi$ and $\Omega$ hyperons and thus further constrain the Hyperon-Nucleon interaction.

\subsection{CHIMERA}
The CHIMERA 4$\pi$ detector was originally conceived in
order to study the processes responsible for particles production in nuclear 
fragmentation, the reaction dynamics and the isospin degree of freedom \cite{pag04}. 
Studies of nuclear EoS in asymmetric nuclear matter have been performed 
both at lower densities with respect to nuclear saturation density, in the Fermi energy 
regime at LNS Catania facilities \cite{def14}, and at high densities in the relativistic heavy ions beams energy domain at GSI \cite{Rus16} (see sec. \ref{sec:asy-eos} for details). 
The possibility to produce, during the dynamics of 
the collision, a short living and transient piece of nuclear matter far 
from ground state properties of temperature, isospin asymmetries and densities is the key 
to extract observables that are sensitive to the symmetry energy components 
of EOS in asymmetric matter and to the in-medium effective interactions of relevance for astrophysical phenomena \cite{hor14, oertel17}. 
From experimental point of view one of the key point of the device 
is the low threshold for simultaneous mass and charge identifications of particles 
and light ions, the velocity measurement by Time-of-Flight technique and the Pulse Shape Detection (PSD) aiming to measure the rise time of signals for charged particles stopping 
in the first Silicon detector layer of the CHIMERA telescopes. Isospin properties of 
Intermediate Mass Fragments (IMF) were used to constrain the density dependence of the symmetry energy at subsaturation densities by using observables based on the transport of isospin 
like the isospin diffusion and N/Z equilibration \cite{sun10,lom10}, the isospin migration of neutrons and protons in the ``neck'' region \cite{def12}. \\
On the other hand, the production of RIB at LNS in the recent years has opened  the use of the 4$\pi$ detector CHIMERA to nuclear structure and clustering studies \cite{acqu16}. 
This requires to add to the charged particle identification techniques the capabilities to 
detect, with good efficiency, also $\gamma$-rays and/or neutrons and to develop new data 
analysis techniques in order to extract high resolution angular distributions from kinematic coincidence measurements. In fact the very high efficiency of the device (4$\pi$ coverage for particle detection and more than 40\% efficiency for $\gamma$-rays detection up to 8 MeV) allows to measure rare coincidences between particles and $\gamma$-rays detected in the second 
detector layer of the CHIMERA telescopes made by CsI(Tl) scintillators. \\
This detection quality has been already demonstrated in various experiments, by using stable and radioactive beams, devoted to the study of pygmy resonances \cite{martorana2018} and on the measurement of the branching ratio for gamma decay of $^{12}C$ states involved in the helium burning phase and in the synthesis of carbon in the Universe \cite{cardella2017}. In this case the 
4$\pi$ setup is exploited by measuring in coincidence $\alpha$ particles (by $\Delta$E-E method), $^{12}$C recoil nucleus (by Time-Of-Flight method) and $\gamma$-rays by pulse shape discrimination in CsI(Tl) scintillators, enhancing the signal to noise ratio in such rare decays. This allows to measure, with good confidence level, also branching ratios around $10^{-7}$ and the study of rare excited levels decay of medium light ions. \\
Moreover, thanks to the availability of intense radioactive beams in the range up to 70 mass units produced by the new FRAISE (FRAgment In-flight SEparator) fragment separator at LNS \cite{RUS18} the systematic study of the excitation and decay of pygmy resonance will be also pursued in order to better assess its importance in the astrophysical environments especially for the r-process. \\
High energy and angular resolution particle-particle correlation measurements are important tools 
in both nuclear structure and nuclear dynamics studies in order to characterize the time scale and shape of emission sources in the dynamical evolution of heavy ion collisions.  FARCOS is an ancillary and compact multi-detector with high angular granularity and energy resolution for the detection of light charged particles and Intermediate Mass Fragments. It has been designed
in order to be coupled with CHIMERA or other devices. The FARCOS array is constituted by 20 telescopes in the final project (complete realization is expected by the end of 2020). Each telescope is composed by three detection stages: the first $\Delta E$ is a 300 $\mu$m thick DSSSD silicon strip detector with 32x32 strips; the second is a DSSSD, 1500 $\mu$m thick with 32x32 strips; the final stage is constituted by 4 CsI(Tl) scintillators, each one 6 cm in length
\cite{epag16}. \\
A further important development for the near future is the capability to add the neutron detection to the list of available observables. In fact this is a critical point in particular when working  with RIB facilities and neutron rich nuclei. A research project has been started aimed to develop the prototype of a  plastic scintillator array based on EJ-276 scintillators in order to detect neutrons in coincidence with light charged particles or IMFs to be coupled or integrated to other devices \cite{epagrad18,epagion18}.

\subsection{GAMMA}
The GAMMA experiment carries out nuclear structure studies via $\gamma$-ray spectroscopy measurements. The activity is based on the use of multi-detector arrays at facilities delivering stable (e.g., LNL Legnaro, GANIL,...) and radioactive (e.g., ISOLDE, GSI/FAIR, RIKEN, ...) ion beams, and intense neutron beams (e.g, ILL-Grenoble), in the framework of international collaborations. A well defined research program is also foreseen at SPES.
The experimental setup always involves the use of high resolution detector arrays, such as the state-of-the-art project AGATA (Advanced Gamma Tracking Array) \cite{AGATA}, and the GALILEO array at LNL. The $\gamma$ spectrometers are most often coupled to complementary detectors, boosting their sensitivity to weak reaction channels, hard to reach with standard $\gamma$-spectroscopy techniques. Notable examples are arrays made of large volume scintillators for high-energy $\gamma$-rays (e.g., the Italian array HECTOR+ of 8 LaBr$_3$(Ce) detectors \cite{GIAZ13} and the international project PARIS \cite{PARIS}), arrays of highly segmented Si detectors for charged particle detection (e.g., the SPIDER \cite{SPIDER} detector and TRACE/MUGAST array \cite{TRACE}), neutron detectors (e.g., the NEDA array \cite{NEDA}), as well as magnetic spectrometers (e.g., PRISMA at LNL and VAMOS at GANIL).\\
Such complex setups will allow to investigate, with unprecedented sensitivity, the structure of astrophysically relevant nuclei. For examples, it will be possible to measure branching ratios for $\gamma$ emission from unbound (resonance) states (via particle-$\gamma$ coincidence techniques), lifetimes of narrow, near threshold resonances (employing Doppler Shift Attenuation techniques down to the femtoseconds time scale \cite{MICHELAGNOLI12}), as well as angular distributions and the electric/magnetic character of the radiation, by using continuous angle variable.

\mbox{} \newline
Key inputs to define reaction rates in stars are: nuclear structure properties of light nuclei relevant for thermonuclear fusion reactions, shapes/deformation and resonance (pygmy) excitation modes of exotic species lying along or in the vicinity of the s- and r-process paths, as well as neutron-capture cross sections and $\beta$-decay half-lives and neutron emission probabilities. As briefly described below, in each case, a significant step forward is expected in the coming years, taking advantage of the combined availability of the multi-detector setups, currently being developed within the GAMMA community, and high-intensity stable and radioactive ion beams:

\begin{itemize}
\item Light nuclei: The structure of light nuclei of Be, Li, C, O, N, ...  plays a key role in thermonuclear fusion reactions in stellar objects. In this context, the measurement of electromagnetic decays from unbound states in neutron-rich systems would represent a breakthrough. At present, such information is very scarce, owing to $\gamma$-decay branchings of the order of 10$^{-4}$-10$^{-3}$ and lower. Of paramount importance are, in particular, near threshold resonances (such as the famous "Hoyle" state in $^{12}$C), most often exhibiting a pronounced cluster structure which directly impacts reaction rates \cite{Oko13,vOer06}.

In the near future, fusion reactions with intense $^{6,7}$Li and $^{14}$C beams on $^{9}$Be, $^{12,13}$C, $^{6,7}$Li and $^{10}$B targets are planned to be employed to populate unbound states in $^{13}$B, $^{15}$C, $^{17,19,20}$O and $^{16,17}$N, after a single evaporated proton, directly feeding the resonance state from which the $\gamma$ decay will be measured.

Astrophysically important proton-rich nuclei involved in the rp-process nucleosynthesis will also be explored at SPES with the available $^{25,26}$Al beams using indirect approaches, such as ($^{3}$He,d) or (d,n) reactions. Knowledge of resonant states in $^{26}$Si and their $\gamma$-decay branch are key to constraint the reaction rates of the $^{25}$Al(p,$\gamma$)$^{26}$Si reaction that are required to  understand the synthesis of $^{26}$Al in explosive hydrogen burning. Thick cryogenic targets will allow a breakthrough in these type of studies:

\item nuclear shapes, deformation and resonance (pygmy) excitation modes: The evolution of nuclear shapes along isotopic chains, including sudden dramatic changes from spherical to deformed systems (as seen in neutron-rich nuclei with Z = 36-40 near N = 60  \cite{Cas09,Tog16,Isk17}), leads to significant changes in $\beta$-decay half lives and proton emission probabilities, with strong impact on the r-process path.
In the near future, Coulomb excitation and cluster transfer \cite{Bot15} reactions with intense exotic beams (in particular neutron rich Kr, Rb, Y and Sr from SPES) will be instrumental for reaching a detailed picture of how nuclear shells evolve away from stability.

Concerning collective excitations, the presence of dipole strength at low excitation energy has been often associated to the oscillation of a neutron skin outside the proton-neutron core: the Pygmy Dipole Resonance mode (\cite{Sav13,Bra15} and reference therein). Its strength is expected to be more pronounced in nuclei far from stability and to play a significant role in the r-process nucleosynthesis \cite{Gor98,Gor04,Roc12} (see Fig. \ref{pygmy1}). Moreover, its properties can be used to constrain the equation of state of neutron rich matter and the radius of low mass neutron stars. At present, scattered information is available for exotic systems of O, Ni and Sn, however, future, extended, systematic investigations are planned with high intensity exotic beams (e.g. along the Sr isotopic chain at SPES) \cite{Lan11}.

\begin{figure}[ht]
\begin{center}
\resizebox{.6\textwidth}{!}{\includegraphics{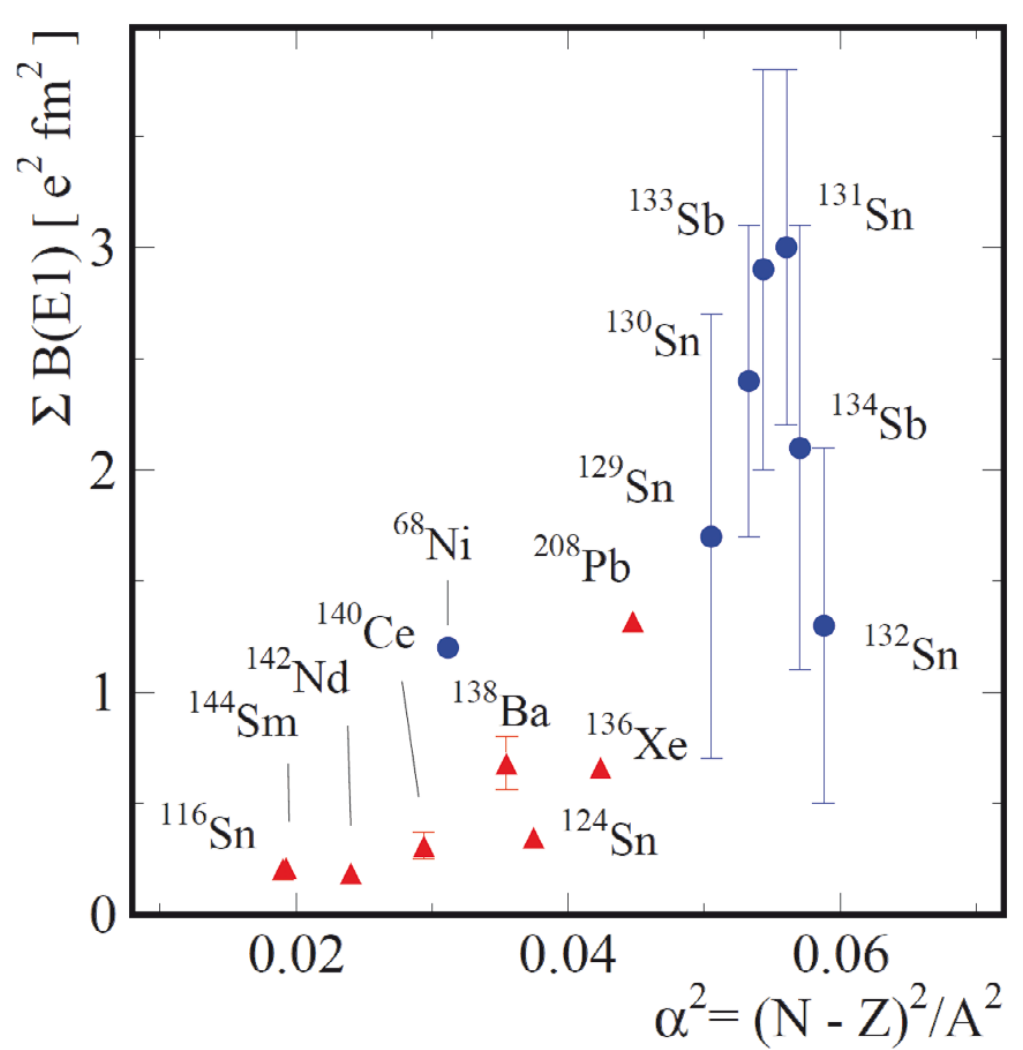}}
\caption{The E1 pygmy dipole strength measured in few nuclei, as a function of the neutron excess \cite{Sav13}.}
\label{pygmy1}
\end{center}
\end{figure}

\item (n,$\gamma$) cross sections: For several astrophysically relevant isotopes, our knowledge of (n,$\gamma$) reaction cross sections is purely theoretical. Some examples on the r-process side are $^{79}$Se, $^{81,85}$Kr, $^{86}$Rb, $^{65}$Zn, $^{121}$Sn. There are also many radioactive nuclei located along or close to the s-process path for which there is neither a theoretical approach nor experimental data. Notable examples are $^{64}$Cu, $^{108}$Ag, $^{109}$Pd, and $^{123}$Sn.

In the near future, neutron capture cross sections for the above mentioned isotopes could be accessed by the measurement of partial cross sections for the population of the final states of the residues formed in the light-ion direct reactions. If the nuclei of interest are exotic, these quantities would be extracted directly via the measurement of light-ion direct reaction in inverse kinematics, performed onto a CD$_2$ target. In the case of stable nuclei, indirect methods could also be used, such as the surrogate method \cite{Esc12} in which the relevant compound nucleus reaction cross section of interest is deduced via a (d,p) reaction study.

\item $\beta$-decay half-lives and neutron emission probabilities: On the neutron rich side, $\beta$ decays, changing neutrons into protons, deviate the r-process path towards heavier species, and their rates define the possible occurrence of successive neutron capture. In this context, key quantities are $\beta$-decay rates, existence of long-living isomeric states and competition with other decay modes, in particular with $\beta$-delayed neutron emission. Such crucial information is fully available for a very limited number of r-process nuclei. Particularly relevant is the region of poorly known neutron rich Ge, As and Se nuclei, which are believed to be responsible for the overproduction of Sr, Y and Zr registered in ultra metal-poor stars.

In the near future, half-lives measurements and the evaluation of probabilities for delayed neutron emission are planned to be pursued at SPES, using the $\beta$-Decay Station setup equipped with plastic and HPGe detectors. Delayed neutron emission measurements would be performed both by studying the internal deexcitation of the successor, as well as by using dedicated setups.
\end{itemize}
\mbox{} \newline
The competitiveness of the astrophysically relevant nuclear structure studies carried out by the GAMMA collaboration resides in the power of the detection setup (e.g., AGATA, GALILEO and their  ancillaries), and in the availability of high intensity stable/radioactive ion beams in Europe, in particular. This makes possible a top class research program in direct competition with similar sophisticated setups, as for example the state-of-the-art $\gamma$ spectrometer GRETA \cite{GRETA}, currently being built in USA. 

\subsection{PANDORA}
The investigation of radioactive decays for astrophysical purposes has become more and more interesting in the last decades: as a matter of fact, not yet solved issues
on stellar and primordial nucleosynthesis require precise radionuclides lifetime measurements.

\mbox{} \newline
The basic idea of PANDORA \cite{pand1} (Plasmas for Astrophysics Nuclear Decays Observation and Radiation for Archaeometry) is that compact
and flexible magnetic plasma traps (plasmas density: $n_e\simeq10^{11}-10^{14} cm^{-3}$, temperature: $T_e\simeq0.1-30 keV$) are the place
where measuring nuclear $\beta$-decay rates at stellar temperatures. This is especially true for radionuclides involved in
nuclear-astrophysics processes and cosmology (s-process, cosmochronometers, Early Solar System formation). The decay rates can be measured as a function of the charge state distribution of the in-plasma ions, and eventually of the plasma density and temperature.
Experiments performed at GSI on highly-charged ions have already demonstrated that the Electron Capture (EC) decay probability can be dramatically modified by the atomic electron configuration. In H-like and He-like $^{140}$Pr,  although the number of atomic orbital electrons was reduced from two to one in $^{140}$Pr$^{57+}$ to $^{140}$Pr$^{58+}$ ions, the EC decay rate increased by a factor of 1.5.

\mbox{} \newline
PANDORA is now in its feasibility study phase, which started in 2017. Two issues are preparatory to nuclear astrophysics experiments: how to measure in-plasma ions charge states distribution (CSD) and how to optimize the handling and the injection of rare isotopes inside the magnetoplasma.\\
Concerning the first issue, a multi-diagnostics characterization of the plasma started at LNS and in other partner laboratories in 2017. The setup includes optical and X-ray spectroscopy,  microwave interferopolarimetry, etc. In addition, the powerful spectropolarimeter SARG (Spettrometro Alta Risoluzione Galileo,  R=160.000 in the range: 370-900 nm) was moved from the Telescopio Nazionale Galileo in Canary Islands to INFN-LNS, where it is currently being installed.
Concerning the second issue, some radionuclides can be handled by conventional vaporization methods. For very short-living species, a Charge Breeding (CB) technique should be applied. As done in ISOL-like facilities such as ISOLDE, SPES and SPIRAL1, 1+ ions produced at the target  can be then injected into a magnetically confined plasma and there in-flight captured. LNS-LNL synergies have permitted to already perform an experiment at LPSC-Grenoble, in late 2017, showing that CB is possible provided that the injection occurs after an ion RF-cooler system.

\mbox{} \newline
Among the various isotopes, some have been selected for their relevance in nuclear astrophysics. A very interesting case is the one of $^{85}$Kr, that is a crucial branching point of the s-process that deeply affects the abundance of Sr and Rb isotopes in stars. Other relevant measurements can be possible for $^{176}$Lu and  the pairs $^{187}$Re-$^{187}$Os and $^{87}$Sr-$^{87}$Rb, which play a crucial role as cosmo-clocks. Lifetimes can be estimated by detecting the emitted $\gamma$-rays. Some of the produced isotopes do not emit $\gamma$-rays. However, alternative ways to estimate the number of decay products can be found, such as their extraction from the plasma and their implantation on special targets that can be then post-processed by XRF techniques to estimate the abundances.
More details on the proposed radionuclides are listed in the following:
\begin{itemize}
\item $^{85}$Kr $\rightarrow$ T$_{1/2}$ = 10.76 y. Involved in s-processes. The decay can be tagged by $\gamma$-rays emission at 514 keV.
\item $^{176}$Lu $\rightarrow$ T$_{1/2}$ = 38 Gy. Used as cosmochronometer. The decay can be tagged by $\gamma$-rays emission at 202.88 and 307 keV
\item $^{187}$Re $\rightarrow$ T$_{1/2}$ = 50 Gy. Used as cosmochronometer. The decay does not produce $\gamma$-rays, but it can be tagged by looking at $^{187}$Os eventually extracted from the plasma.
\item $^{87}$Rb $\rightarrow$ T$_{1/2}$ = 48 Gy, (cosmochronometers). The decay does not produce $\gamma$-rays, but it can be tagged by looking at $^{87}$Sr eventually extracted from the plasma.
\end{itemize}

\mbox{} \newline
Nuclear decay lifetimes of radionuclides of astrophysical interest have never been measured in magnetized plasmas. The measurements that are planned in PANDORA will be competing/complementing the ones already performed on Storage Rings (SR) (see for instance \cite{pand2}). The main advantage with respect to SR is that in PANDORA even long-lifetime isotopes can be studied, since the magnetoplasma can be tuned in a dynamical equilibrium (i.e. keeping constant the plasma density, temperature and charge state distribution) for days, or even weeks.

\mbox{} \newline
The interest for PANDORA in groups working worldwide in the field of ECR plasma traps is constantly growing. As a matter of fact, GANIL and LPSC-Grenoble (in France), Jyvaskyla University (in Finland) and the ATOMKI laboratory of the Hungarian Academy of Sciences (in this case, an official letter of interest has been endorsed) have already presented informal or even formal letters of intents for starting a scientific collaboration. As mentioned before, the measurements of some of the radionuclide lifetimes is now a topic of some storage ring facilities, such as the one in the design phase at HIE-ISOLDE.

\subsection{SPES}

The SPES radioactive ion beam facility, presently under construction al LNL, is based on the isotope separation on-line technique.  A primary proton beam, provided by the driver accelerator, is used to induce nuclear reactions inside a thick target like silicon carbide or uranium carbide. The reaction products are extracted from the target by thermal processes due to the high temperature of the target-ion-source system. Once reached the source the reaction products are ionized ($1^+$) and extracted. After an isotopic selection the radioactive elements undergo a subsequent ionisation (n$^+$ charge state) by means of a charge breeder to be further re-accelerated by the LNL superconductive linear accelerator ALPI. Accelerated radioactive ions, both in the low energy branch (after isotopic selection) or in the high energy branch (after passing the linac), are then used for basic nuclear research and for nuclear physics based applications. The primary proton beam is delivered by a cyclotron accelerator with an energy of 40 MeV and a beam current of 200 $\mu$A. Isotopic selection is obtained by the High Resolution Magnetic Spectrometer (HRMS) coupled to a Beam Cooler (BC). To boost the charge state of the produced radioactive ions, and allow post acceleration by a newly designed Radio-Frequency Quadrupole (RFQ) coupled to the existing superconductive linear accelerator ALPI, an Electron Cyclotron Resonance (ECR) based charge breeder (CB) has been realized. To suppress contaminants introduced by the ECR a Medium Resolution Magnetic Spectrometer (MRMS) has been added to the beam line. The aim is to provide re-accelerated high intensity (I = $10^7$- $10^9$ particle per second) mainly neutron-rich secondary beams at energies of about 10 MeV/A, an energy regime allowing to overcome the Coulomb barrier to all stable targets.

Main aim of the proposed research is the study of the reactions that are at the basis of the element formation in the universe.  Heavy elements from iron up to uranium are produced by neutron captures by the r- and s-processes. The study of those neutron capture reactions, essential for a quantitative understanding of the element formation in the universe, is the main objective of the astrophysical research at SPES.

\mbox{} \newline
Since nuclei involved in such processes are generally short-living and a neutron target cannot be realized, such reaction studies have mainly to rely on indirect methods. The experimental techniques proposed are the Trojan Horse (THM) and the Surrogate Ratio (SRM) methods.
The SRM uses a different (surrogate) reaction to produce the compound nucleus of interest which, by definite kinematical conditions, populates the same excited states \cite{Potelprc,Potelepja}. The method assumes that the formation of the compound nucleus can be accurately described by nuclear models. The decay of the compound nucleus, where different channels are competing with each other as a function of neutron energy, is the goal of the experimental investigation. The measured decay probabilities of the compound nucleus are then used to tune model parameters leading to the neutron-induced cross sections. Suitable reactions are exchange reactions with light nuclei, e.g. (d,p). In such a case a radioactive ion beam is directed onto a deuterium gas target. During the particle exchange the neutron of the deuterium target is passed on to the projectile nucleus and the remaining proton is detected. The knowledge of the full kinematics of the proton allows to select the excitation energy level of the compound. Such excited compound nucleus will then de-excite via $\gamma$-emission (capture), neutron emission (inelastic scattering) etc. The measurement of the $\gamma$ decay of the compound will allow to select the different channels.
Another case of interest is the inelastic-scattering reaction of deuterons - denoted (d,d$'$) - where the deuteron is scattered leaving the heavy projectile in an excited state.
Very important is the influence of the populated angular momentum and parity on the decay probabilities. In most cases the differences between the distributions populated by the surrogate and the desired neutron-induced reaction have to be taken into account through a normalization to a reference case (surrogate ratio method).

\mbox{} \newline
Future nuclear astrophysics measurements will involve:
\begin{itemize}
\item s-process nucleosynthesis: Capture cross sections of the nuclei only produced by s-process are essential for testing the models that describe the formation of heavy elements. Among the long list of branching point isotopes we shortly discuss two examples: $^{85}$Kr(n,$\gamma$) reaction rate relevant for the study of the $^{85}$Kr branching point of the s-process. The neutron capture cross section on $^{85}$Kr, competing with the beta decay of this unstable isotope with T$_{1/2} \approx$11 y, is sensitive to the neutron density in the environment of the AGB star where the s-process takes place. \\
$^{79}$Se(n,$\gamma$): Elements with masses equal or heavier than A=80 are supposed to be synthesized by the s-process operating in AGB stars, where two alternate burning shells, one of H and an inner shell of He, surround an inert degenerate CO core. The He inter-shell is periodically swept by convective instabilities induced by He-burning runaway where $^{12}$C is synthesized. The main neutron source for AGB stars of mass M $\leq$ 4 solar masses and T $\approx10^{8}$ K is the $^{13}$C($\alpha$,n)$^{16}$O reaction (main s-process component) with a neutron density of N$_n$ $\leq10^7$ cm$^{-3}$. In the case of more massive stars, the s-process takes place during pre-supernova evolution, i.e. during convective core He-burning and convective C-shell burning. In this case also the neutron source $^{22}$Ne($\alpha$, n) $^{25}$Mg  is activated at temperatures of T $\geq 3.5 \cdot 10^8$ K and  neutron densities of the order of N$_n$ $\geq$ 10$^{10}$ cm$^{-3}$ can be reached.  Different s-process patterns are expected depending on whether the first or the second neutron source reaction is more active. The knowledge of the neutron capture cross sections in such branching nuclei together with the observed element abundances allow to estimate the neutron density at the s-process site.

\item r-process nucleosynthesis: Final r-process abundances are sensitive to neutron capture cross sections during freeze-out. The SPES neutron-rich beams are essential for testing reference neutron capture rates, probing the stability of the nuclear models used to extrapolate the r-process nucleosynthesis path.
Examples of capture reactions to be investigated with the early phase SPES beams are:
$^{81}$Ge(n,$\gamma$), $^{78,80,81}$Ga(n,$\gamma$), $^{129,131}$Sn(n,$\gamma$),
$^{134}$Sb(n,$\gamma$), $^{132,134,136,138}$Te(n,$\gamma$), $^{137}$Xe(n,$\gamma$).

\item Light nuclei direct reactions with light beams from SPES: Using the SiC target, light beams can be provided by the SPES radioactive ion beam facility. Examples of charged particle reactions of interest for the physics of novae and of X-ray bursts that can be performed at SPES are listed below:\\
Novae: \\
$^{18}$F(p,$\alpha$)$^{15}$O, $^{30}$P(p,$\gamma$)$^{31}$S, $^{25}$Al(p,$\gamma$)$^{26}$Si,$^{26}$Al(p,$\alpha$)$^{23}$Mg using THM\\
X-ray burst:\\
$^{14}$O($\alpha$,p)$^{17}$F, $^{18}$Ne($\alpha$,p)$^{21}$Na, $^{30}$S($\alpha$,p)$^{33}$Cl.
\end{itemize}

\mbox{} \newline
Most of the proposed reactions are based on beams not available at existing facilities with sufficient intensities.

\mbox{} \newline
Main competition in Europe is CERN-ISOLDE (CH). In the future FRIB (USA) and the upgrade of TRIUMF (Canada) will provide more intense secondary beams. It is highly important to keep the time schedule of SPES in order to not spoil the discovery potential of the facility.
\markright{EoS of ultra-dense matter}

\section{Equation of State of dense and ultra-dense matter}
In a {\it classical} neutron star model, the core includes the most internal
99\% of the total mass and consists of superfluid neutrons and superconducting protons, plus electrons and muons.
However, if the density exceeds about $6 \cdot 10^{14}$ g/cm$^3$, more exotic material
may arise. Various possibilities have been suggested, among which: meson condensates ($\pi$, K), hyperons (strange baryons) and quarks. In the latter case the neutron star
is called {\it hybrid}. All these components would modify the equation of state (EoS),
 thus affecting
 the mass-to-radius relation. When a neutron star belongs to a binary system,
both masses and radii can be measured. These measurements provide an ultimate constraint to
the ultra-high density EoS. In particular, the maximum mass of a stable neutron star
(the Tolman-Oppenheimer-Volkoff limit) is smaller in case of exotic matter. Latest measurements
of neutron star masses in binaries seems to exclude hybrid models (for a review see \cite{lattimer2016}).
In all cases, to correctly interpret these observations,
a more reliable theory describing the thermodynamic properties of
ultra-high density matter is absolutely required. For a comprehensive review of the neutron star EoS see \cite{Piekarewicz-2016}.

\mbox{} \newline
Hyperon-nucleon and hyperon-nucleon-nucleon interactions are studied by ALICE in connection
to the modeling of astrophysical objects like neutron stars.
As a matter of fact, the appearance of hyperons in the inner core of these objects
is often energetically favored but it is also responsible for a softening of the EoS that
makes the EoS itself incompatible with the observation of two neutron stars with mass as large
as 2 solar masses.
This puzzle might be solved by models which introduce three-body forces and which need
to be constrained by a detailed knowledge of the hyperon-nucleon interaction and of the
hyperon-nucleon-nucleon interaction. These interactions will
be further studied in several experiments involving Italian groups. In particular,
ALICE will study hyperon-nucleon interactions and the production of hypernuclei
during LHC Run-3 and Run-4, while complementary studies of hypernuclear spectroscopy
will be performed at the Jefferson Laboratory (JLab). In addition, the SIDDHARTA-2 (Silicon Drift Detector for Hadronic Atom Research by Timing Application) and the AMADEUS (Antikaonic Matter At DAFNE: Experiments with Unraveling Spectroscopy) collaborations
will investigate kaon-nucleon and kaon-nuclei interactions at the DAFNE collider, and
the ULYSSES (UnraveLing hYpernuclear Structure and Spectroscopy ExperimentS) experiment at the Japanese facility J-PARC will try to constrain the three-body forces in ($\Lambda$-)hypernuclei. Finally, ASY-EOS (ASYmmetric-matter Equation-Of-State) will explore in the high density region (2.5-3 $\rho_{0}$), the EoS and the Symmetry Energy term, by studying the elliptic flow in heavy ion collisions. 

\markright{Experiments on the Equation of State of ultra-dense matter}
\section{Experiments on the EoS of dense and ultra-dense matter}

\subsection{ALICE} \label{alice-eos}
\begin{figure}[!h]
  \centering
   \includegraphics[width=0.8\textwidth]{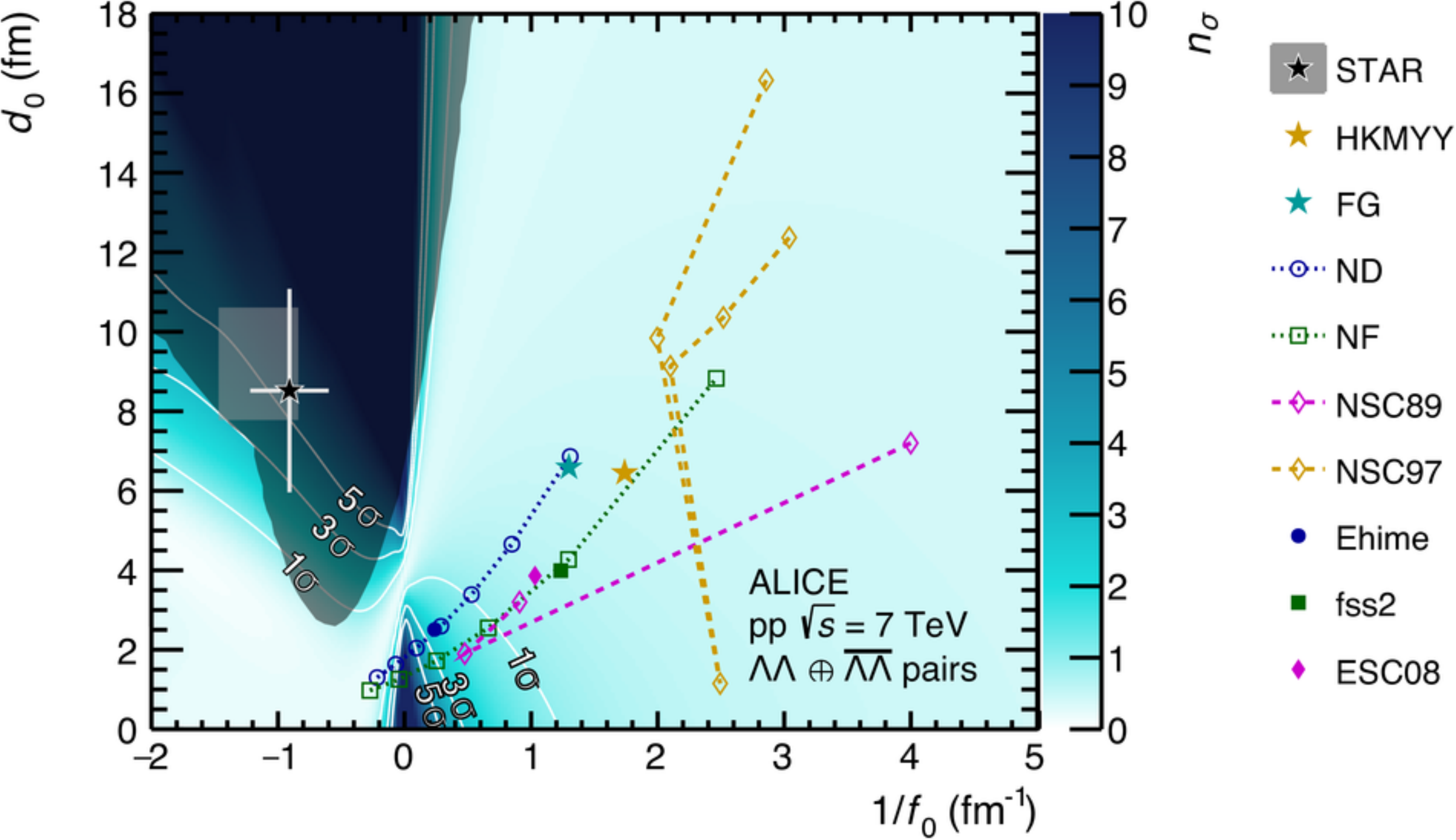}
  \caption{Number of standard deviations of the modeled $\Lambda$-$\Lambda$  correlation  function for a given set of scattering parameters (effective range $d_0$ and scattering length $f_0$) with respect to the ALICE experimental data~\cite{Acharya:2018gyz}.}
\label{alice1}
\end{figure}

The study of the hyperon-nucleon and hyperon-nucleon-nucleon interaction has become more and more
crucial in recent years due to its connection to the modeling of astrophysical objects like neutron
stars. In the inner core of these objects the appearance of hyperons is a probable scenario,
because their formation is often energetically favored in comparison with a purely nucleonic matter
composition. However, the appearance of these additional degrees of freedom leads to a softening of
the matter equation of state (EoS) that makes the EoS incompatible with the observation of two neutron
stars of two solar masses~\cite{Demorest:2010,Antoniadis:2013}.
This leads to the `hyperon puzzle'~\cite{Bombaci:2017}. Many attempts were made to solve this puzzle,
e.g.\,by introducing three-body forces leading to an additional repulsion that can counterbalance the large gravitational pressure and finally allow for large star masses. To constrain the parameter space of such models a detailed knowledge of the hyperon-nucleon interaction (HNI)  and of the hyperon-nucleon-nucleon interaction is mandatory, including $\Lambda$, $\Sigma$ and $\Xi$ states.\\
The hyperon-nucleon (YN) interaction can be studied using femtoscopy, i.e.\,the correlation function of particle pairs at low relative momentum, in particular in small colliding systems as pp, where the emitting
source is well constrained (e.g. by precise proton-proton femtoscopy).
A first ALICE measurement~\cite{Acharya:2018gyz} demonstrates that the p-$\Lambda$ and $\Lambda$-$\Lambda$ correlation functions constrain the phase space spanned by the effective range and scattering length of the strong interaction. The results for the $\Lambda$--$\Lambda$ case are summarized in Fig.~\ref{alice1}.

\begin{figure}[!h]
  \centering
   \includegraphics[width=0.7\textwidth]{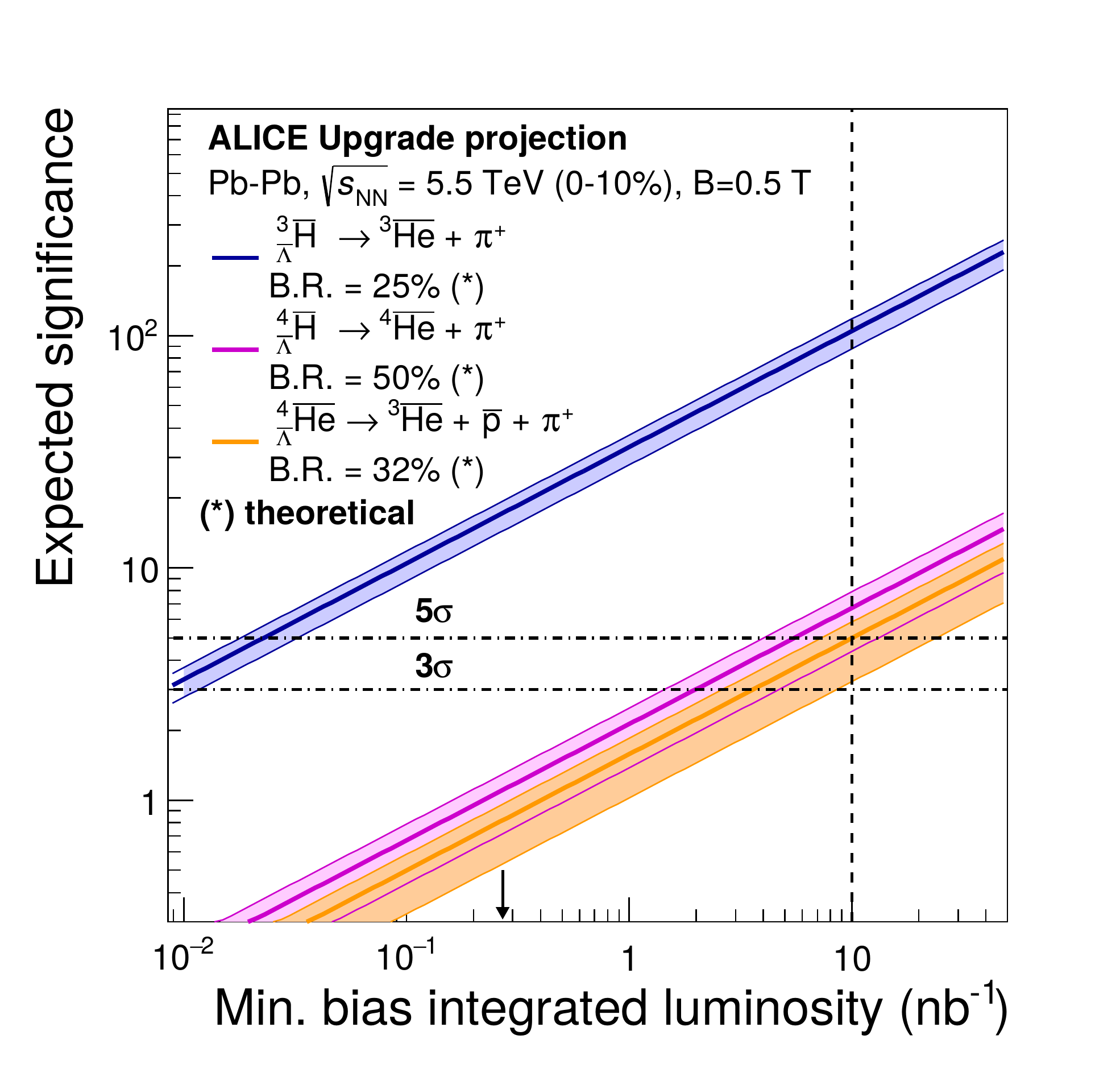}
  \caption{Projected significance of anti-hyper-nuclei measurements in central Pb--Pb collisions in Runs-3 and 4 as a function of the integrated minimum-bias luminosity. The arrow represents the minimum bias Pb--Pb luminosity anticipated for the end of Run 2. The vertical line represents the ALICE target. The bands indicate the uncertainty on the significance associated with different model predictions~\cite{Andronic:2010qu, Bellini:2018epz}.}
\label{alice2}
\end{figure}

\mbox{} \newline
A hypernucleus is a nucleus that contains at least a hyperon in addition to protons and neutrons. The lifetime of a hypernucleus depends on the strength of the hyperon-nucleon (YN) interaction. The study of this interaction is relevant for nuclear physics and nuclear astrophysics. For example, it plays a key role in understanding the structure of neutron stars. Depending on the strength of the YN interaction, the collapsed stellar core could consist of hyperons, strange quark matter, or a kaonic condensate.
The Pb-Pb data set collected by ALICE in the LHC Run-1 and Run-2 already allowed for a first yield and lifetime measurements
of hypertriton $^3_{\Lambda}$H and anti-hypertriton~\cite{Adam:2015yta}.
The detection of heavier (anti)-hypernuclei will be possible with future Run-3 and Run-4 samples.\\
The data sample of Pb--Pb collisions that will be collected during LHC Run-3 and Run-4 will be larger by a factor 100 with respect to that of the ongoing Run-2.
This sample will enable very precise measurements of the production of $^3_{\Lambda}$H and first ever experimental measurement  of
$^4_{\Lambda}$H and $^4_{\Lambda}$He via their weak decays, see Fig. \ref{alice2}.

\subsection{ASY-EOS}
\label{sec:asy-eos}
Outer core of a neutron star consists of superfluid neutrons and superconducting protons, together with electrons and muons, in a density phase above nuclear saturation desity $\rho_{0}$. As told before, main properties of neutron stars are governed by EOS and, in particular, by Symmetry Energy $S({\rho})$. In fact, the density dependence of the Symmetry Energy $S({\rho})$ above $\rho_o$ can be accessed by the determination of the masses and radii of neutron stars \cite{Lat14}. Instead, in a terrestrial experiment, high-density dependence of $S({\rho})$ can be investigated by means of heavy-ion collisions, reproducing in a laboratory nuclear matter above saturation density,  by employing observables related to the early-high density phase of the reactions.  A multitude of observables have been proposed to be sensitive to $S({\rho})$ at supra-saturation densities (for a review see \cite{Bao08}): ratio of multiplicities or spectra of isospin partners (e.g. $\pi^-/\pi^+$, n/p, t/$^3$He, K$^{+}$/K$^{0}$). The ratio of positively and negatively charged pions measured close to or below the production threshold in the NN system ($\sim280$~MeV) is one of the observables mainly discussed, but, in the past,  controversial results have been obtained \cite{xiao,feng,danielewicz}. It is expected that more reliable results will be obtained by measuring pion yield ratio as a function of kinetic energy \cite{Spirit}.\\
Other observables which are sensitive to $S({\rho})$ at supra-normal densities  are collective flows. At energies below 1 GeV per nucleon the reaction dynamics is largely determined by the nuclear mean field. The resulting pressure produces a collective motion of the compressed material whose strength will be influenced by the $S({\rho})$ in Isospin asymmetric systems.  The strengths of
collective flows in heavy ion collisions are determined by a Fourier expansion of the azimuthal distributions of particles around the reaction plane: $d\sigma(y)/d\phi \propto 1 + 2(v_1(y) \cos{\phi} + v_2(y) \cos{2\phi}.....)$, where $y$ is the rapidity and $\phi$ the azimuthal angle with respect to the reaction plane \cite{And06}. The side flow of particles is characterized by the coefficient $v_1$ and the elliptic flow by $v_2$. The value of $v_2$ around mid-rapidity is negative at incident beam energies between 0.2 and 6 GeV per nucleon which signifies that matter is squeezed out perpendicular to the reaction plane. Elliptic flow at those energies is known to be sensitive to the stiffness of the symmetric part of the nuclear EOS \cite{lefevre, rei12}. Instead, for the case of Isospin asymmetric matter, the ratio
$v_{2}^n/v_2^{p}$ of elliptic flow strengths of neutrons to that of protons -or Hydrogens (H), or light charged particles (LCP)- was recommended in \cite{Rus11} as a robust observable sensitive to the stiffness of  $S({\rho})$.
The $v_{2}^n/v_2^{H}$ was  measured in Au+Au semi-peripheral collisions at 400 AMeV, studied at GSI laboratory by means of FOPI charged particle detectors and LAND (Large Area Neutron Detector) neutron/LCP detectors. The comparison of the measured observable with transport calculations based on Ultra relativistic Quantum Molecular Dynamics - transport model (UrQMD) model has allowed to give a first constraint for $S({\rho})$ behavior above ${\rho_0}$, with the slope
parameter $L = 83 \pm 26$~MeV \cite{Rus11}. Despite a large uncertainty, the result made it possible to rule out the extremely soft or stiff density dependencies of symmetry energy. This result has been confirmed by using a different transport code \cite{Coz13}. A more recent experiment, carried out at GSI laboratory in 2011, by coupling several devices (CHIMERA, AToF-Wall, MicroBALL, Kratta, LAND) was able to better measure $v_{2}^n/v_2^{LCP}$ in Au+Au collisions at 400 AMeV; the obtained value of the slope parameter is $L$=72$\pm$13~MeV, improving by a factor 2 the accuracy on the L value with respect to the former study \cite{Rus16}. A careful study carried out with T\"ubingen QMD calculations allowed to establish that density region explored in those measurements lies mainly between 0.8 and 1.6 $\rho_0$. Thus, ASY-EOS experimental results proved the effectiveness of the elliptic flow ratio in constraining the high-density behavior of $S({\rho})$. The obtained result is in agreement with ref. \cite{Zha18}, extracting $S({\rho})$ from observations of neutron stars and gravitational waves.\\
Next step will be to extend this kind of measurements toward larger incident energy, up to about 1-1.5 AGeV, capable of producing nuclear matter up to about 2.5-3 ${\rho_0}$, also by profiting of improved capabilities of new devices, as the NeuLAND detectors now in operation at the GSI laboratory \cite{AELOI}. Moreover, the measurement of  of the $v_{2}^n/v_2^{p}$ in a wide energy range may allow the determination of the $K_{Sym}$, the curvature of $S({\rho})$ around $\rho_{0}$, that governs considerably $S({\rho})$ behavior when moving further away from ${\rho}_0$ and that, up to know, is substantially unconstrained \cite{Cozlast}.

\subsection{JLab}
\begin{figure}[h]
 \centering
   \includegraphics[width=0.65\textwidth]{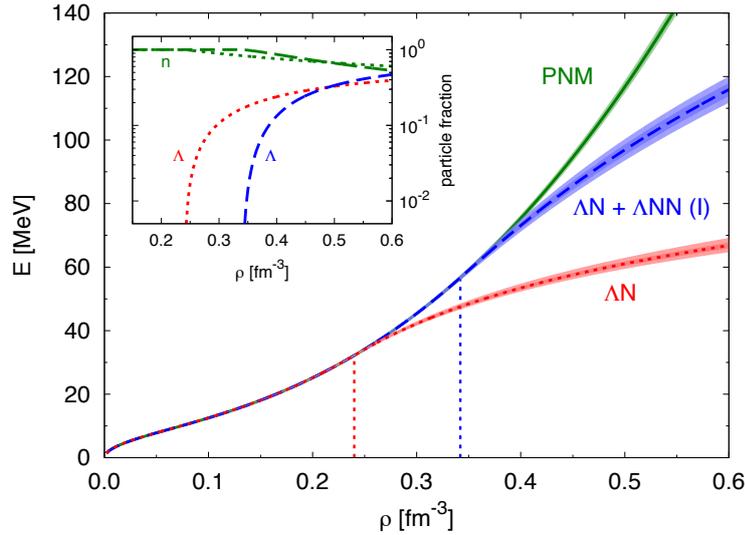}
  \caption{Equations of state. The vertical dotted lines indicate the $ \Lambda $ threshold
densities. In the inset neutron and $ \Lambda $ fractions correspond to the two hyper-neutron matter EOSs (from \cite{Lonardoni-2015}).}
\label{jlab1.4}
\end{figure}

\begin{figure}[h]
 \centering
      \includegraphics[width=0.65\textwidth]{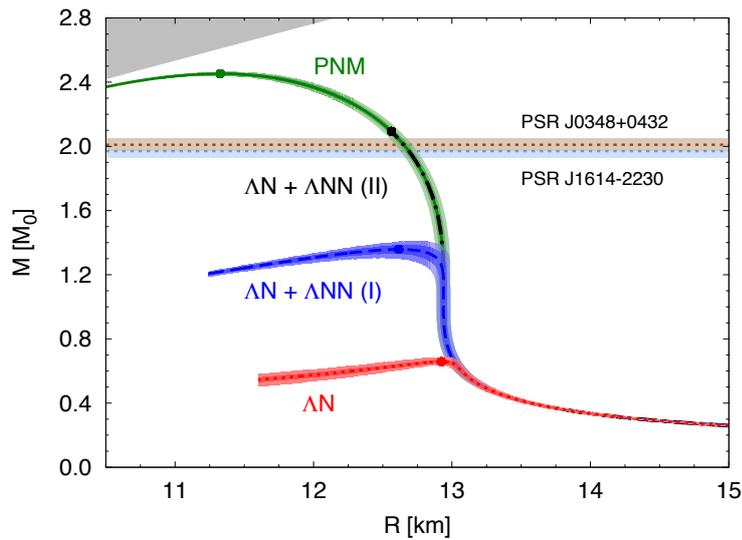}
  \caption{Mass-radius relations as calculated by the Auxiliary Field Diffusion Monte Carlo
(AFDMC). Closed circles represent the predicted maximum masses. Horizontal bands at 2 solar masses are the observed masses of the heavy neutron stars (from \cite{Lonardoni-2015}).}
\label{jlab1.5}
\end{figure}

The appearance of hyperons in the inner core of neutron stars is thought to be energetically favored.
In fact, in the pure neutron matter, whenever the chemical potential becomes sufficiently large to match
the chemical potential of a hyperon in the same matter, the hyperon becomes stable since it is a
distinguishable particle, and creates its own Fermi sea, thereby lowering the kinetic energy of
the system.
However, the presence of hyperons in the inner core of the neutron stars is also responsible for a
softening of the EoS that makes the EoS itself incompatible with the observation of two neutron
stars with mass as large as two solar masses. This puzzle, called hyperon-puzzle, might be solved
by models which introduce repulsive three-body forces at high densities.
A hint for the introduction of this kind of forces can be derived from the non-strange nuclear sector:
 when only two-body forces are accounted for, light nuclei turn out to be under-bound and the
 saturation properties of infinite isospin-symmetric nuclear matter are not correctly reproduced.
 This indicates the need for a three-body interaction.\\
 The binding energies of light nuclei have been used to constrain three-nucleon potential models.
 However, the most accurate phenomenological three-body force (Illinois 7), while providing a
 satisfactory description of the spectrum of light nuclei up to $^{12}$C,
 leads to a pathological EoS
 for pure neutron matter (PNM). On the other hand, when additional information on the three-nucleon
 interaction is inferred from saturation properties of symmetric nuclear matter,
 the resulting PNM EoS turns out to be stiff enough to be compatible with astrophysical observations.
 Recent analysis of $^{16}$O-$^{16}$O scattering data shows that the established meson exchange
 potential model (Nijmegen ESC08c) cannot reproduce the cross section at large scattering angles
 and inclusion of 3-body/4-body repulsive forces solves the problem.
 Thus, there is a general indication that 3-body/4-body repulsive forces become quite significant
 at high density in the non-strange nuclear sector, and one can reasonably suppose that the same
 applies in the strange nuclear sector. To confirm this hypothesis accurate measurements determining
 with great precision hyperon-nucleon and hyperon-nucleon-nucleon interactions are needed.
 Due to the present unavailability of hyperon beam facility, and hence the impossibility of
 collecting hyperon-neutron and hyperon-hyperon scattering data,
 the only ways so far to derive realistic hypernuclear interaction models rely on information
 extracted from the binding energies of hypernuclei and from the Pb-Pb collisions.
The latter will be collected during LHC Run-3 and Run-4 (see section \ref{alice-eos}),
while a hypernuclear spectroscopy program
is under development at JLab.
As regards the hypernuclear spectroscopy, in analogy with the
non-strange nuclear sector where, as quoted above,
3-body/4-body repulsive forces cannot be constrained from light systems, the binding energies of
light hypernuclei do not suffice in constraining hypernuclear interactions.
Additional information must hence be inferred from the properties of medium and heavy hypernuclei
in order to extrapolate to the infinite-mass limit.\\
The inclusion of the three-body $\Lambda NN$ force leads to a satisfactory description of the hyperon separation energies in a wide mass range of hypernuclei and for the $\Lambda$ occupying different
single particle state orbitals (s, p and d wave).
However, these potential models predicting relatively small differences in the $\Lambda$ separation
energies of hypernuclei give dramatically different results as for the properties of the infinite medium.
The resulting EoS spans the whole regime extending from the appearance of a substantial fraction
of hyperons at $\sim 2\rho_0 \sim 0.32$ fm$^{-3}$ to the absence of $\Lambda$ particles in the entire
density range of the star (see Fig. \ref{jlab1.4}), with consequent sizable effect on the predicted
neutron stars structure (see Fig. \ref{jlab1.5}).

Very accurate measurements of the $\Lambda$ separation energy
of medium to heavy hypernuclei are hence of fundamental importance to determine hyperon-nucleon
and hyperon-nucleon-nucleon interactions with the precision needed to understand neutron star
features in general and to solve, in particular,  the hyperon puzzle. JLab PAC has already approved
one experiment of hypernuclear spectroscopy aiming to determine the isospin dependence of the
$\Lambda N$ interaction, that plays a fundamental role in neutron star structure. Since neutron stars are made by more than 90\% of neutrons the effects of isospin asymmetry (that one can translate into an isospin dependence of the hyperon nucleon force) are very important.
The experiment will measure the $\Lambda$ separation energy of the two hypernuclei: $^{40}_{\Lambda}$K
, obtained through the reaction $^{40}$Ca(e,e'K$^{+})^{40}_{\Lambda}$K and very little sensitive
to isospin asymmetry contributions having a nearly equal number of neutrons and protons (for symmetric
hypernuclei the Pauli principle suppresses any strong contribution from the $\Lambda nn$ or $\Lambda pp$
channels) and $^{48}_{\Lambda}$K, obtained through the reaction $^{48}$Ca(e,e'K$^{+})^{48}_{\Lambda}$K,
much more sensitive to contributions arising from $\Lambda nn$ triplets (for asymmetric hypernuclei
the contribution of the channel with the two nucleons in isospin triplet (T = 1) state is enhanced.
This feature is maximum in neutron matter or in matter at $\beta$-equilibrium and hence in neutron stars).\\

Another experiment that will have important implications on our knowledge of neutron star features is PREX (Pb Radius EXperiment). PREX is measuring the neutron skin, i.e. the difference between the neutron and the proton radius in $ ^{208}$Pb. This information is extracted from the Parity Violating Asymmetry (APV), i.e. the fractional difference in cross sections for positive and negative helicity electrons elastically scattered off $^{208}$Pb. The measurement of $ R_N $, the $^{208}$Pb neutron radius, constrains the EoS of neutron matter. There is a strong correlation between $ R_N $ and the pressure of neutron matter P: a larger P would push neutrons out, in contrast with the surface tension, and it will increase $ R_N $. A larger $ R_N $ implies a stiffer EoS with a larger pressure that, in turn, would imply larger neutron star radii.  In addition, the EoS of neutron matter is connected to the symmetry energy S, which is related to the nuclear matter energy change as function of the difference between the number of neutrons and the number of protons. A large S, at high density, would mean a large proton fraction in neutron stars. If a big difference between the neutron and the proton radius in $ ^{208}$Pb is measured, then it is likely that massive neutron stars cool quickly by direct Urca process producing neutrinos freely leaving neutron stars \cite{Urca}. In addition, the transition density from the solid neutron star crust to the liquid interior is strongly correlated with the difference between the neutron and the proton radius. A big difference (stiffer EoS) would imply a thin crust (and vice versa) \cite{Crust}.  PREX has already collected a first set of data \cite{PREX}. A new set of data that will increase the statistics will be collected this year. 

\subsection{SIDDHARTA-2 and AMADEUS Collaborations at DAFNE}
Kaon-nucleon and kaon-nuclei interactions at low-energies, studied by SIDDHARTA-2 and AMADEUS
Collaborations at the DAFNE Collider, using the best quality low-energy kaon beam in the world,
provide fundamental information for constraining the EoS for neutron stars, especially regarding the
role that strangeness may play in the core of a neutron star. SIDDHARTA-2 is measuring kaonic atoms
 which deliver the isospin-dependent antikaon-nucleon scattering lengths, i.e. the interaction at
 zero relative energy between kaons and nucleons. AMADEUS is studying the interactions of low-energy
 kaons (below 100 MeV/c) with nuclear matter,  resolving for the first time branching ratios and
 cross sections in final channels containing hyperons, where multi-hadron interaction processes are
 resolved.  These data are used in all QCD models at
low-energy (chiral theories, potential models, few-body calculations etc.) and, as such,
are fundamental in resolving between various EoS for neutron stars.\\
SIDDHARTA-2 will measure for the first time the kaonic deuterium in 2019-2020; future measurements of
 other light and heavier kaonic atoms, to characterize the potential for the interaction of kaons
  with nuclei, are being proposed after this run. AMADEUS is finalizing the analyses of data
  in various channels coming from the interaction of kaons with nuclear matter containing strangeness,
   while preparing a dedicated proposal for future studies of selected process, especially important
    for the EoS of neutron stars.

\subsection{ULYSSES at J-PARC}
The most natural and suitable environment where to investigate the properties of the three-body forces are the $\Lambda$-hypernuclei, which will be systematically studied in the framework of the ULYSSES experiment at the Japanese facility J-PARC.
Thanks to its so called glue-like role, the addition of the $\Lambda$ hyperon to an unstable nucleus leads to the creation of a stable bound system.
It is then possible to observe neutron-rich $\Lambda$-hypernuclei, like for instance $_\Lambda^6$H \cite{FIN12A,FIN12B,JE1014,JE1017} and $_\Lambda^9$He \cite{FIN12C}, that are appreciably beyond the neutron stability drip line in nuclear systems.
Their study could place valuable constraints on the size of coherent $\Lambda N - \Sigma N$ mixing in dense strange neutron-rich matter \cite{AKAI99,SHIN02,AKAI08,AKAI10}.
Such a mixing provides a robust mechanism for generating three-body $\Lambda NN$ interactions, with a direct impact on the stiffness or the softness of the equation of state for hyperons in neutron-star matter \cite{SHAF08,SHAF10}.
Strangeness plays another important role in neutron stars as far as their cooling is concerned.
The presence of hyperons provides an additional mechanism that makes the cooling process faster thanks to, for instance, neutrino emission through the
$Y \rightarrow B + l + \bar \nu_l$ reaction.
However, hyperon superfluidity tends to suppress the faster cooling. So the attention is focused on the quantitative estimation of hyperon pairing.
In the new generation of gravitational wave observatories in the world, the gravitational wave emitted during the NS-NS merger is supposed to be a good indicator of the source. Numerical simulations of the NS-NS merger are carried out based on the EoS of the neutron star.
These studies suggest that the gravitational wave pattern could be sensitive to the neutron star EoS \cite{AFTN15}. 

\markright{Conclusion}
\section{Conclusions}
In the report we discussed at first the scenarios where nuclear astrophysics plays a key role, shortly describing the related INFN activities. Then, we gave rather detailed status and prospects  of both the experiments essentially devoted to nuclear astrophysics and of a few experiments which are also contributing to nuclear astrophysics. Finally, we describe the inputs from several different experiments to better constrain the equation of state of dense and ultra-dense matter.
It is clear that Italian groups financed by INFN are giving
quite important contributions to every domain of nuclear astrophysics, having sometimes been the initiators of worldwide unique experiments.
For instance, fundamental inputs have been provided for an accurate calculation of the solar neutrino spectrum, for the prediction of BBN isotope abundances and for the firm identification of the stellar environments where a few key isotopes have been synthesized. We have also seen that important issues are studied with different but complementary approaches by different groups, this way reaching sound and comprehensive solutions to several astrophysical problems. The future of experimental nuclear astrophysics within INFN appears very promising. As a matter of fact, in addition to the natural evolution of the ongoing activities, new beams will soon become available at three INFN national laboratories provided by SPES at Legnaro, by the 3.5 MV accelerator under Gran Sasso and by the forthcoming noble gas and long-lived radioactive isotope beams at LNS.

\clearpage

\section*{Acknowledgments}
It is a pleasure to thank all the Colleagues who contributed to achieve the results discussed in this article.

\markright{References}
\printbibliography

\end{document}